\documentclass[twocolumn, prl, superscriptaddress,amsmath,amssymb,aps,pra,prb,prstab,prstper]{revtex4}
\usepackage{amsmath}
\usepackage{subfigure}
\usepackage{graphicx}
\usepackage{dcolumn}
\usepackage{bm}
\usepackage[normalem]{ulem}
\usepackage{amssymb}
\usepackage{soul}
\usepackage{xcolor}
\usepackage{xparse}
\usepackage{color}
\usepackage{enumitem}
\usepackage{enumitem}
\usepackage{pgfplots}
\pgfplotsset{compat=1.9}
\usepackage{tikz}
\usepackage{hyperref}

\begin{document}

\preprint{APS/123-QED}

\title{Bright traveling breathers in media with long-range, nonconvex dispersion}

\author{Sathyanarayanan Chandramouli}
\email{sathyanaraya@umass.edu}
 \affiliation{Department of Mathematics and Statistics, University of Massachusetts at Amherst.}
\author{Yifeng Mao}%
 \email{yifeng.mao@colorado.edu}
\affiliation{Department of Applied Mathematics, University of Colorado Boulder}%
\author{Mark A. Hoefer}
\email{hoefer@colorado.edu}
\affiliation{Department of Applied Mathematics, University of Colorado Boulder}%
\date{\today}

\begin{abstract}
  The existence and properties of envelope solitary waves on a
  periodic, traveling wave background, called traveling breathers, are
  investigated numerically in representative nonlocal dispersive
  media.  Using a fixed point computational scheme, a space-time
  boundary value problem for bright traveling breather solutions is
  solved for the weakly nonlinear Benjamin-Bona-Mahony equation, a
  nonlocal, regularized shallow water wave model, and the strongly
  nonlinear conduit equation, a nonlocal model of viscous core-annular
  flows.  Curves of unit-mean traveling breather solutions within a
  three-dimensional parameter space are obtained.  Resonance due to
  nonconvex, rational linear dispersion leads to a nonzero oscillatory
  background upon which traveling breathers propagate.  These
  solutions exhibit a topological phase jump, so act as defects within
  the periodic background.  For small amplitudes, traveling breathers
  are well-approximated by bright soliton solutions of the nonlinear
  Schr{\"o}dinger equation with a negligibly small periodic
  background.  These solutions are numerically continued into the
  large amplitude regime as elevation defects on cnoidal or
  cnoidal-like periodic traveling wave backgrounds.  This study of
  bright traveling breathers provides insight into systems with
  nonconvex, nonlocal dispersion that occur in a variety of media such
  as internal oceanic waves subject to rotation and short, intense
  optical pulses.
\end{abstract}


\maketitle


\section{\label{sec:level1}INTRODUCTION\protect\\}
Coherently propagating disturbances such as solitary waves and
envelope solitary waves, ubiquitous in nonlinear dispersive media, are
formed due to a balance between nonlinearity and dispersion.  Envelope
solitary waves are generically described by solutions of the cubic
nonlinear Schr\"odinger (NLS) equation subject to two disparate
spatial scales: the fast spatial scale corresponding to the wavelength
of periodic carrier wave oscillations and the slow-amplitude-phase-modulation scale.  For an attractive, focusing nonlinear medium, the
NLS equation admits bright soliton solutions corresponding to a
localized, sech envelope modulation of a rapidly varying carrier wave.
When the carrier phase speed and envelope group speed differ, NLS
bright solitons approximate unsteady nonlinear wave packets.  Such
nonlinear wavepackets are also referred to as breathers due to the
pulsation or breathing of their internal oscillations.  However, the mere
existence of approximate NLS bright soliton solutions does not
guarantee the existence of breather solutions to a nonlinear
dispersive evolution equation.

The canonical evolution equations admitting breather solutions are the
sine-Gordon (SG) \cite{ablowitz_solitons_1981} and modified
Korteweg-de Vries (mKdV) \cite{clarke2000generation} equations.  These
equations are completely integrable by the inverse scattering
transform where soliton and breather solutions correspond to discrete
eigenvalues of a corresponding linear spectral problem
\cite{ablowitz_solitons_1981}.  A breather solution of SG or mKdV equations
corresponds to a {pair} or quartet of eigenvalues, {respectively},
which can be interpreted as a bound state of two solitons that decay
to zero \cite{clarke2000generation}.  Breather solutions to the
focusing NLS equation have also been identified as important for
modeling rogue wave phenomena \cite{onorato_rogue_2013} and exhibit a
nonzero, plane wave background.

The concept of breathers as interacting soliton pairs has been
generalized to solutions in which a soliton interacts with a periodic
traveling wave, e.g., a cnoidal wave, using a variety of exact
solution methods that are available for integrable systems
\cite{kuznetsov_stability_1975,hu2012explicit,cheng2014interactions,ren2015interaction,chen_periodic_2019,pelinovsky_localized_2020,bertola_partial_2022,hoefer_kdv_2023}.
Such solutions have been interpreted as dislocations of a cnoidal
background \cite{kuznetsov_stability_1975}.  We refer to these
solutions as {traveling breathers}, which is consistent with
their interpretation in lattice systems
\cite{flach_discrete_2008,james2018travelling}.  Traveling breathers
are spatially localized on a periodic traveling wave background and
they exhibit two distinct velocities, the phase velocity of the cnoidal
background and the envelope velocity of the traveling breather.  Other
terminology that has been used to describe these coherent structures
includes quasi-breather \cite{ablowitz_solitons_1981}, localized
oscillatory state \cite{campbell_resonance_1983}, and nonlocal
solitary wave \cite{boyd_numerical_1990}.  

{In the context of interfacial waves that arise in a viscous
  core-annular flow, coherent bright breather trains were observed to
  form in computational runs of perturbed, modulationally unstable
  periodic waves \cite{maiden2016modulations}. The nonlinear
  Schrödinger equation was utilized as an approximate framework to
  study these wavepackets in the weakly nonlinear regime. A follow-up
  work, involving extensive experiments, generated bright and dark
  traveling breathers at the interface of such viscous core-annular
  flows by interacting solitons and cnoidal-like waves
  \cite{mao2023observation}. These traveling breathers were seen to
  robustly persist within the experimental test section for 15--25
  oscillatory periods and long distances.}

{Such breathers and
traveling breathers} are also prevalent, for example, in fluid dynamics
\cite{lamb_breather_2007,grimshaw2008long,whitfield2015wave,maiden2016modulations,nakayama2020breathers},
nonlinear and matter-wave optics
\cite{leblond2013models,hasegawa1995solitons,kivshar1998dark,mandelik_observation_2003,schafer2004propagation,sakovich2005short,di_carli_excitation_2019,luo_creation_2020},
and magnetic materials
\cite{kalinikos1990spin,ustinov2008observation,drozdovskii2010formation}.
However, due to their inherently unsteady character, breather and
traveling breather solutions are challenging to obtain. {As noted earlier, the exact breather waveforms for integrable systems are constructed through a nonlinear ``superposition" principle based on an application of the Darboux transformation. On the other hand, a common approach to circumvent the reduced analytical tractability associated with nonintegrable systems} involves long-time numerical evolution of suitably
chosen initial conditions that appear to lead to breather solutions
\cite{helfrich2007decay,lamb_breather_2007,grimshaw2008long,dyachenko2008formation,whitfield2015wave,maiden2016modulations,nakayama2020breathers}.
However, due to the existence of small-amplitude radiation accompanying
the time evolution, it is difficult to discern whether breather
solutions actually exist and, if so, what their properties are.  The
existence of radiation in these and other computational studies
\cite{kudryavtsev_solitonlike_1975,ablowitz_solitary_1979,campbell_resonance_1983}
suggests that a more likely scenario for nonintegrable systems is
that breathers are accompanied by an oscillatory background
\cite{segur_nonexistence_1987,boyd_numerical_1990,soffer_resonances_1999-1},
recent numerical evidence of localized breathers in a nonintegrable
equation notwithstanding \cite{kalisch2022breather}.  In other words,
just as solitary-wave solutions in nonintegrable equations generalize
the soliton solutions of integrable equations, traveling breather
solutions of nonintegrable equations are the natural generalization
of breather solutions of integrable equations.

In the present study, we numerically investigate the existence of
bright (elevation) traveling breathers to the Benjamin-Bona-Mahony
(BBM) and conduit equations, both nonintegrable, nonevolutionary
equations \cite{mikhailov_classification_2007}, by solving a
space-time boundary-value problem (BVP) in the comoving reference
frame where the envelope speed is zero. {The boundary conditions are periodicity in time and space.} Multiple one-dimensional
families of traveling breather solutions are obtained by numerically
continuing the BVP solutions from the weakly nonlinear Schr\"odinger
bright soliton approximation with a given carrier wave number and amplitude.  The
unit-mean carrier frequency, phase shift, and amplitude are implicitly
determined by fixing the breather velocity, the carrier frequency in the
comoving frame, and the spatial domain size.  Solution branches are
obtained by performing continuation in the traveling breather velocity,
which is negative for all solution branches computed.

Solutions along a given continuation branch are found to strongly
depend upon the initial carrier wave number.  When the wave number is
sufficiently far from the inflection point of the linear dispersion
relation, the carrier background amplitude grows with decreasing velocity
while the traveling breather width narrows relative to the carrier
wavelength.  For initial carrier wave numbers close to the inflection
point, the traveling breather envelope width remains large relative to
the carrier wavelength.  The traveling breather solutions obtained
here are found to be dynamically stable under long-time numerical
evolution subject to small-amplitude initial noise. {Since nonlinear
short-pulse optics \cite{costanzino2009solitary,leblond2013models} and internal oceanic
waves influenced by the earth's rotation
\cite{helfrich2007decay,whitfield2015wave} exhibit similar nonconvex
rational dispersion, the traveling breathers obtained in our study may
have implications for these and other applications.}

We adapt the Newton-conjugate gradient (NCG) method
\cite{yang2009newton,yang2015numerical} to compute traveling breather
solutions. Our parametric continuation scheme is detailed in
Sec.~\ref{c-continuation}. It is essential to seed the iterative
continuation scheme with good initial guesses.  We initialize the NCG
iterations with weakly nonlinear NLS approximations described in
Sec.~\ref{Computational-methodology}. 

\section{Model equation properties }

The BBM equation in normalized form \cite{peregrine_1966,benjamin1972model}
\begin{equation}
  \begin{aligned}
    \label{BBM-equation-1}
    &u_t+uu_x-u_{xxt}=0,
  \end{aligned}
\end{equation}
is a long-wavelength model of weakly nonlinear waves.  Equation
\eqref{BBM-equation-1} is not integrable, possessing exactly two other
linearly independent conservation laws \cite{olver_1979}. Besides the
usual space and time translational-invariance properties, the BBM
possesses the scaling symmetry
\begin{equation}
   \label{Scaling-symmetry-bbm-eq}
   u\rightarrow u_0 u,\hspace{1mm}x\rightarrow x,\hspace{1mm}t\rightarrow u_0t,
\end{equation}
where $u_0$ is a nonzero real constant.  The BBM equation's linear
dispersion relation for trigonometric traveling waves on the
constant background $u_0$ is bounded 
\begin{equation}
   \label{BBM-Dispersion-relation}
   { \omega_0({k},u_0) = \frac{u_0 {k}}{1+{k}^2}}
\end{equation}
and exhibits zero dispersion when $u_0=0$ or ${k}=\sqrt{3}$ since
\begin{equation}
    \label{BBM-GVD}
    \partial_{{k}{k}}
    \omega_0({k},u_0) =
    \frac{2u_0{k}({k}^2-3)}{({k}^2+1)^3}.
\end{equation}
The bounded, nonconvex dispersion \eqref{BBM-Dispersion-relation}
distinguishes the short-wave behavior of BBM solutions from those of
the Korteweg-de Vries (KdV) equation $u_t + uu_x + u_{xxx} = 0$ with
unbounded dispersion and no inflection points for nonzero $k$.

The BBM equation \eqref{BBM-equation-1} admits a three-parameter
family of periodic traveling-wave solutions in the form of cnoidal
waves
\begin{equation}
  \label{eq:7}
  \begin{split}
    u(x,t) &= \tilde{\beta} + (\tilde{\gamma} - \tilde{\beta}) \mathrm{cn}^2 \left ( z, m \right
             ), \\
    z &= \left (\frac{\tilde{\gamma} - \tilde{\alpha}}{12 \tilde{s}} \right )^{1/2} (x - \tilde{s} t),
        \quad m = \frac{\tilde{\gamma} - \tilde{\beta}}{\tilde{\gamma} - \tilde{\alpha}} ,
  \end{split}
\end{equation}
where $\tilde{\alpha} < \tilde{\beta} < \tilde{\gamma}$,
$\tilde{s} = \frac{1}{3}(\tilde{\alpha} + \tilde{\beta} + \tilde{\gamma})$ is the phase velocity and
$\mathrm{cn}(z,m)$ is the Jacobi elliptic cosine function.  The
cnoidal wave's amplitude $a$ and wavenumber $k$ are
\begin{equation}
  \label{eq:9}
  a = \tilde{\gamma} - \tilde{\beta}, \quad
  k = \frac{2\pi}{L}, \quad L = 4 K(m) \sqrt{\frac{3\tilde{s}}{\tilde{\gamma} -
      \tilde{\alpha}}} ,
\end{equation}
while its mean is 
\begin{equation}
  \label{eq:8}
  \overline{u} = \tilde{\alpha} + (\tilde{\gamma} - \tilde{\alpha}) \frac{E(m)}{K(m)} ,
\end{equation}
where $K(m)$ and $E(m)$ are the complete elliptic integrals of the
first and second kinds, respectively.  The cnoidal wave \eqref{eq:7}
limits to a solitary wave when $\tilde{\beta} \to \tilde{\alpha}$ and a constant when
$\tilde{\beta} \to \tilde{\gamma}$.

By use of the scaling symmetry \eqref{Scaling-symmetry-bbm-eq}, we
impose the unit-mean constraint $\overline{u} = 1$ on the cnoidal wave
solutions without loss of generality and therefore constrain
$\Tilde{\alpha}$, $\Tilde{\beta}$, and $\Tilde{\gamma}$ via $\overline{u} =1$ in
eq.~\eqref{eq:8}.  We parametrize the set of unit mean periodic
traveling-wave solutions to the BBM equation in terms of two
parameters such as $(a,k)$.  Then its frequency is determined to be
$\omega = k\Tilde{s}$.

A strongly nonlinear generalization of the BBM equation is the conduit
equation \cite{olson1986solitary}
\begin{equation}
  \begin{aligned}
    \label{Conduit-equation-1}
    &A_t+2AA_z-AA_{tzz}+A_tA_{zz}=0,
  \end{aligned}
\end{equation}
modeling large amplitude, long waves along the circular, free
interface between two viscous fluids with high viscosity contrast and
small Reynolds number \cite{lowman2013dispersive}.  Its linear
dispersion relation on the constant background $A_0 > 0$ is similar to the
BBM equation \eqref{BBM-Dispersion-relation}
\begin{equation}
   \label{Conduit-equation-disp-relation}
    \omega_0({k};A_0) = \frac{2 A_0{k}}{1+A_0{k}^2} .
\end{equation}
It is bounded and has an inflection point when
${k} = \sqrt{3/A_0}$.  We mention that, like the BBM equation,
the conduit equation is not integrable and possesses at least two
independent conservation laws \cite{harris1996conservation}. Finally,
the equation satisfies the scaling symmetry
\begin{equation}
  \label{Scaling-symmetry-conduit-eq}
  A\rightarrow A_0 A,\quad z\rightarrow
  A_0^{-1/2}z,\quad t\rightarrow A_0^{1/2}t.
\end{equation}

The conduit equation \eqref{Conduit-equation-1} also admits a
three-parameter family of periodic traveling-wave solutions
\cite{maiden2016modulations,johnson_modulational_2020} but an
analytical expression for it is not known.  Utilizing the scaling
symmetry \eqref{Scaling-symmetry-conduit-eq}, we impose the unit-mean
constraint so that the cnoidal-like periodic traveling-wave solutions
are parametrized by, e.g., their wave number $k$ and amplitude $a$.

\section{Computational methodology}
\label{Computational-methodology}

We begin by introducing the common approach to computing approximate
bright breather solutions by assuming weak nonlinearity and scale
separation.  The NLS equation models the slowly varying envelope
{$B(\zeta,\tilde{\tau})$} of nearly monochromatic nonlinear wavepackets
\cite{ablowitz2011nonlinear,whitham2011linear}.  {In this regard,
  the focusing NLS reduction obtained by employing a standard multiple-scales calculation and rescaling takes the form
\begin{equation}
  \label{BBM-HNLS}
  \begin{split}
    iB_{\tilde{\tau}}+\dfrac{1}{2}B_{\zeta\zeta}+|B|^2B=0.
  \end{split}
\end{equation}
The coordinate system associated with eq.~\eqref{BBM-HNLS} is
\begin{equation}
\label{Slow-coordinate-system}
  \zeta=\dfrac{\epsilon}{\sqrt{\partial_{\Tilde{k}\Tilde{k}}\omega_0}}(x
  -\partial_{\Tilde{k}} \omega_0t), \quad \tilde{\tau}={\epsilon^2}t, 
\end{equation}
where $\partial_{\Tilde{k}}\omega_0$ is the group velocity,
$\partial_{\Tilde{k}\Tilde{k}}\omega_0$ represents the dispersion
curvature, and $\epsilon$ is an amplitude scale. The benefit of the NLS
approximation is that a simple ordinary differential equation (ODE)
can be sought for describing the spatial variation of the envelope
$b(\zeta)$, where $B=b(\zeta)\exp(i\mu \tilde{\tau})$ and $\mu$ is an
amplitude-dependent frequency shift. The well-known sech solution of
this ODE, $b\equiv {\rm sech}(\zeta)$ and $\mu=\frac{1}{2}$ will be used to seed the
continuation algorithm that we describe shortly.}

In contrast, the direct computation of traveling breathers requires
solving a partial differential equation.  We now describe the
strategy we adopt to compute BBM and conduit bright traveling
breathers as solutions to a space-time boundary-value problem.

\subsection{Space-time boundary value problem}
\label{sec:space-time-boundary}
The BBM and conduit equations are examples of nonevolutionary
equations in the form
\begin{equation}
  \label{prototype-pde}
  u_t=\mathbb{L}[u,u_t]+\mathbb{N}[u,u_t],
\end{equation}
where $\mathbb{L}$ is a linear, constant coefficient skew-adjoint
differential operator while $\mathbb{N}[u,u_t]$ is in general a
nonlinear operator acting on $u$ and $u_t$.  Entering the comoving
frame with velocity $c$ $(\chi=x-ct,\tau =t)$, we recast
eq.~\eqref{prototype-pde} as
\begin{equation}
\begin{aligned}
  \beta_{\tau} - c\beta_{\chi}-\tilde{\mathbb{L}}[1+\beta,\beta_{\tau}
                -
                c\beta_{\chi}]-\tilde{\mathbb{N}}[1+\beta,\beta_{\tau}
                - c \beta_{\chi}] = 0,
    \end{aligned}
    \label{Space-time-BVP-unit-background}
\end{equation}
where $u(x,t) \equiv 1+\beta(\chi,\tau)$ and the linear operator
$\Tilde{\mathbb{L}}$ inherits the skew symmetry.  Using the scaling
symmetry \eqref{Scaling-symmetry-bbm-eq} or
\eqref{Scaling-symmetry-conduit-eq}, we set the background mean to
unity, without loss of generality.  In this reference frame, the
solution is assumed to be time periodic with period $T$ and to rapidly
decay to a periodic background in space.  Then $\beta$ has zero mean
in the far field in $\chi$ due to the unit-mean normalization of $u$.
We truncate the domain $(\chi,\tau) \in [-L,L]\times [0,T]$ and take a
finite Fourier product basis for the solution field
\begin{equation}
\begin{aligned}
\beta(\chi,\tau;c,T)&\approx\sum_{m=-N}^{N} \alpha_m(\chi) e^{im({2\pi \tau}/{T})},\\
\alpha_{m}(\chi)&\approx\sum_{s=-M}^{M} \hat{\alpha}_{ms}
                  e^{is({2\pi \chi}/{L})} .
\end{aligned}
\label{Product-basis}
\end{equation}
The spatial domain is chosen to be sufficiently large so that boundary
effects are negligible, leading to a fully periodic product basis. The
actions of the two-dimensional forward and inverse discrete Fourier
transforms are denoted by $\mathcal{F}_{2{\rm D}}(\cdot)$ and
$\mathcal{F}_{2{\rm D}}^{-1}(\cdot)$, respectively.

{We now describe the iterative procedure to recover numerical
  solutions from their weakly nonlinear approximations governed by the
  NLS equation, which in {the co-traveling frame, and in terms of fast space and time variables is }
  {\begin{align}
\label{Internal-dynamics-BBM}
  &\beta(\chi,\tau)\approx \\\nonumber& {\frac{\tilde{a}}{2}}\; {\rm
  sech} \left( \frac{\epsilon}{\sqrt{\partial_{\Tilde{k}\Tilde{k}}\omega_0}} {\chi}
  \right) \cos \Big( \tilde{k} \chi- (\omega_0-\frac{\epsilon^2}{2}
  -\tilde{k} \partial_{\Tilde{k}}\omega_0)\tau\Big),
\end{align}
where {$\epsilon = \frac{\tilde{a}\sqrt{\gamma}}{4}$} while the
parameters $\gamma(\tilde{k})$ and $\omega_0$ depend on the particular dispersive evolution equation at hand.} Having fixed the time period $T$ to the NLS
  prediction, the solution field $\beta$ and the envelope velocity $c$
  need to be determined.}  As we will show, the family of unit-mean
solutions is three dimensional.  While the ansatz
\eqref{Product-basis} for $\beta$ can be used in
Eq.~\eqref{Space-time-BVP-unit-background}, to determine the velocity
$c$, we require an additional condition.  Multiplying
Eq.~\eqref{Space-time-BVP-unit-background} by $\beta_{\tau}$ and
isolating the terms containing the velocity $c$ results in
$\mathcal{H}(\beta)=c\mathcal{G}(\beta)$.  Integrating this expression
over the entire spatiotemporal domain, we obtain the self-consistent integral condition for $c$,
\begin{equation}
  \label{Self-consistency}
  c =
  \frac{\int_{0}^{T}\int_{-L}^{L}\mathcal{H}(\beta)d{\chi}d{\tau}}
  {\int_{0}^{T}\int_{-L}^{L} \mathcal{G}(\beta) d{\chi}d{\tau}} ,
\end{equation}
provided the denominator is nonzero.  Other integral relations for $c$
may similarly be derived.  Equation \eqref{Self-consistency} is used
because it is found to be robust in the sense that iterations
converge, it is efficient in the associated iterative procedure, and
we never observe the denominator to go to zero.

We implement the Newton-conjugate gradient algorithm on
eq.~\eqref{Space-time-BVP-unit-background} subject to the ansatz
\eqref{Product-basis}.  To recover $c$ and $\beta$ simultaneously, we
update the velocity iteratively at every outer Newton iteration using
\eqref{Self-consistency} (see \cite{yang2015numerical} for a similar
treatment).  Upon insertion of \eqref{Product-basis} into
\eqref{Space-time-BVP-unit-background}, Newton operator iterations are
then applied to Eq.~(\ref{Space-time-BVP-unit-background}).  The
linearization is symmetrized and inner preconditioned-conjugate
gradient iterations are used to solve this positive-semidefinite
self-adjoint system, which is expected to converge
\cite{yang2009newton}. The complete velocity-solution profile update
algorithm is summarized as
\begin{equation}
  \begin{aligned} 
    {\textbf{P}}_{1n}^{\dagger}{\textbf{P}}_{1n} \Delta \beta
    &= -{\textbf{P}}_{1n}^{\dagger}{\textbf{P}}_0\beta_{n},\;\;\;
      n=1,2,_{\cdots}\\
    c_n
    &=\frac{\int_{0}^{T}\int_{-L}^{L}\mathcal{H}_{n}(\beta)d{\chi}d{\tau}}
      {\int_{0}^{T}\int_{-L}^{L} \mathcal{G}_{n}(\beta) d{\chi}d{\tau}},\\
    \Delta \beta &= \beta_{n+1}-\beta_{n},
  \end{aligned}
  \label{NCG-iteration-final-BBM}
\end{equation}
where ${\textbf{P}}_{1n}$ is the linearization operator at the
$n^{}$th Newton iteration, with its adjoint operator represented by
${\textbf{P}}_{1n}^{\dagger}$, and ${\bf P}_0\beta_{n}$ is the
residual of Eq.~\eqref{Space-time-BVP-unit-background} at the $n^{}$th
iteration.

We seed the iterations with a sufficiently close initial guess
{($\beta_1 \;{\rm and}\; c_1$)}. For the BBM and conduit equations, the bright
solitary-wave solutions of the corresponding {NLS} reductions give
sufficiently close approximations.  Iterations are terminated when
{the residual $\underset{\chi,\tau}{\rm max} \;|{\bf P}_0 \beta|$ is
  less than $10^{-10}$.} To reduce the condition number of the linear
operator ${\textbf{P}}_{1n}^{\dagger}{\textbf{P}}_{1n}$, it is
necessary to introduce an acceleration operator. We follow the
guidelines outlined in \cite{yang2009newton,yang2015numerical}.  For
Eq.~(\ref{Space-time-BVP-unit-background}), an appropriate
acceleration operator at the $n^{}$th Newton iteration is chosen by
examining the constant-coefficient part of the symmetrized operator
$ {\textbf{P}}_{1n}^{\dagger}{\textbf{P}}_{1n}$, given by
$-(\mathbb{B}_n)^2=-(\partial_{\tau}-c_n\partial_{\chi}-\tilde{\mathbb{L}})^2$. Notably,
this operator is positive semidefinite.  An appropriate acceleration
operator is thus given by the positive-definite operator
${\bf M}_n=r-(\mathbb{B}_n)^2$, where $r>0$ is a positive number whose choice is arrived at via numerical experiments.
 
\subsection{Parametrization and numerical continuation}
\label{c-continuation}
{Having described the computation of a numerically accurate
  waveform in the weakly nonlinear regime, we now lay out the
  continuation procedure to obtain a family of solutions from this
  known solution.}  Implicit to the product basis representation in
Eq.~\eqref{Product-basis} for spatially localized unit-mean breathers,
we have a two-parameter characterization (envelope velocity $c$ and
timeperiod $T$). Given a known solution, we perform a line search for
a fixed timeperiod $T$, by varying the velocity $c$ parameter
(referred to as $c$ continuation).  We perform several such line
searches starting from different {weakly nonlinear} solutions.
The search algorithm at the $j^{}$th $c$-continuation step is
\begin{equation}
  \begin{aligned}
    \label{NCG-continuation-strategy}
    {\textbf{P}}_{1n}^{\dagger(j)}{\textbf{P}}_{1n}^{(j)} \Delta \beta&= -{\textbf{P}}_{1n}^{\dagger(j)}{\textbf{P}}_0^{(j)}\beta_{n}; n=1,2,_{\cdots},
  \end{aligned}
\end{equation}
{where the iterations are seeded with a known solution ($\beta^{(j-1)}$ and $c^{(j-1)}$), while a traveling breather with fixed velocity $c^{(j)}=c^{(j-1)}+\delta c$ is sought.}
Note that similar space-time boundary-value computations have been
used to compute breathers and traveling breathers in discrete lattice
systems \cite{james2018travelling} and modified NLS-type models
\cite{ward2019evaluating,ward2020rogue}.

\section{Results}
\label{sec:results}

\subsection{BBM bright traveling breather solutions}
\label{sec:bbm-bright-traveling}

We compute {five} branches of BBM traveling breathers, bifurcating
from the focusing NLS limit. The edge of the focusing regime [$\sigma > 0$ in
Eq.~\eqref{BBM-HNLS}] of the NLS reduction is marked by the inflection
point of the unit-mean linear dispersion relation
\eqref{BBM-Dispersion-relation} $\tilde{k}=\sqrt{3} \approx 1.73$.  It
is crucial to initialize the Newton iterations with an accurate
initial guess in the nearly monochromatic regime for convergence to
traveling breathers. In order to use a NLS bright soliton as the
initial seed for the continuation procedure, it is necessary to
initialize the carrier wavenumber $\tilde{k} > \sqrt{3}$.  The
{five} traveling breather branches are characterized by the
carrier wave numbers
\begin{equation}
  \label{eq:1}
  \tilde{k}^{(1)} \approx
  5.42 > \tilde{k}^{(2,3)} \approx 3.79 > \tilde{k}^{(4)}\approx
  3.01 > \tilde{k}^{(5)} \approx 2.42 .
\end{equation}
We refer to each of these branches as families 1, 2, 3, 4, and 5,
respectively. {We perform the computations on large domains for each of the breather families 1-5, with
$2L \in \{400,200,500,500,500\}$, respectively}. {The spatial
discretization for all the computed families is $\Delta x={2L}/{2M}=0.08$, while $N=16$ (32 Fourier modes) is found to be an appropriate discretization in time.} The NLS approximations provide good
initial guesses for computing traveling breathers with carrier wave
numbers $\tilde{k}>2.4$.

{The initial guess for NCG iterations is the
bright soliton solution of the NLS equation~\eqref{BBM-HNLS}, where {$\gamma(\tilde{k})=(5\Tilde{k}^2+3)/(6\Tilde{k}^3+18\Tilde{k})$} and
$\omega_0$ [Eq.~\eqref{BBM-Dispersion-relation}], $\partial_{\Tilde{k}}\omega_0$ and
$\partial_{\Tilde{k}\Tilde{k}}\omega_0$ are evaluated at
$\tilde{k} = \tilde{k}^{(j)}$ for some $j$ and $u_0=1$.}  We
initialize the computations with {$\tilde{a} \approx 0.15 $}.  The NLS
bright soliton envelope is localized, whereas the computed traveling
breathers are found to exhibit oscillatory tails, albeit very small
for small $\tilde{a}$. It is convenient to introduce the traveling
breather frequency in the comoving frame $\Omega = 2\pi/T$.

\begin{figure}
  \centering
  \includegraphics[width=\linewidth]{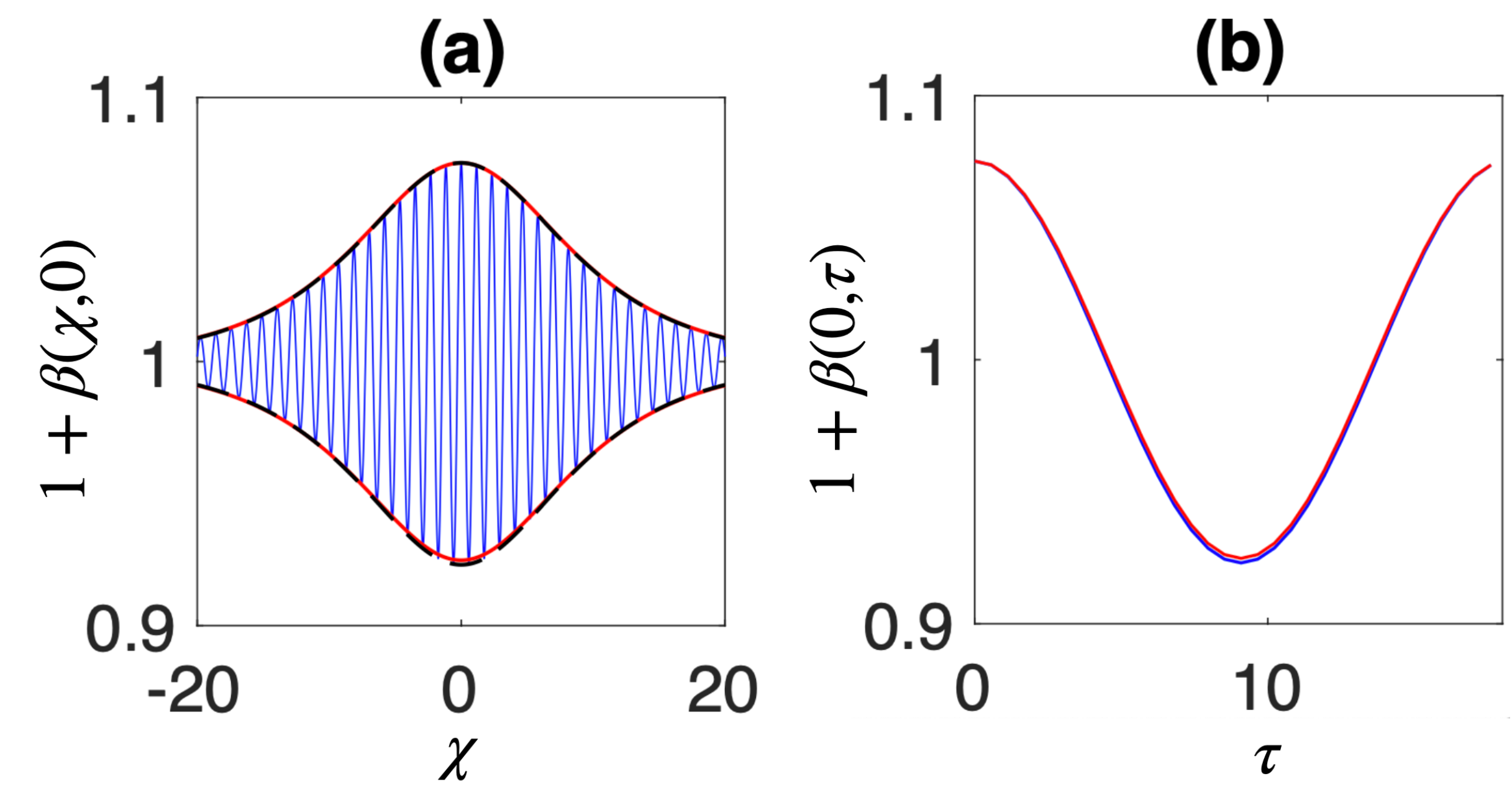}
  \caption{(a) Traveling breather in the weakly nonlinear regime
    (blue solid line) of family 1 with {$T \approx 18.180$\;{\rm and}\; $c \approx -0.031 $}
    on a spatial domain with $L = 400$.  The numerical envelope
    (black dashed line) is compared to the NLS bright soliton envelope
    (red solid line). (b) Evaluation of the traveling
    breather at $\chi = 0$ (blue solid line) compared with the leading-order
    NLS approximation~~\eqref{Internal-dynamics-BBM} (red solid line).}
  \label{fig_BBM_NLS_comp}
\end{figure}
In Fig.~\ref{fig_BBM_NLS_comp} we compare the computed and NLS bright
soliton profiles for the first traveling breather of {family 1}. The profiles agree to within $1\%$. The numerical
envelope profiles are extracted using the Hilbert transform
\cite{nakayama2020breathers}. Furthermore, the traveling breather
frequency $\Omega$ and velocity $c$ are within $1\%$ of the bright
soliton predictions \eqref{Internal-dynamics-BBM}.
The first computed solution for all traveling breather families
exhibits similar agreement to the NLS bright soliton prediction
\eqref{Internal-dynamics-BBM}.

To recover other traveling breathers, we apply $c$ continuation. The
continuation procedure is found to introduce a shift in the wave mean.
%
At the end of $c$ continuation for each family, we scale all the
obtained traveling breathers to unit mean by utilizing
Eq.~\eqref{Scaling-symmetry-bbm-eq}. Consequently, although we fix
the time period $T$ during continuation, the rescaling to unit mean
implies that the time period of the computed traveling breathers
changes across each family.

\begin{figure}
    \centering
    \includegraphics[width=0.8\linewidth]{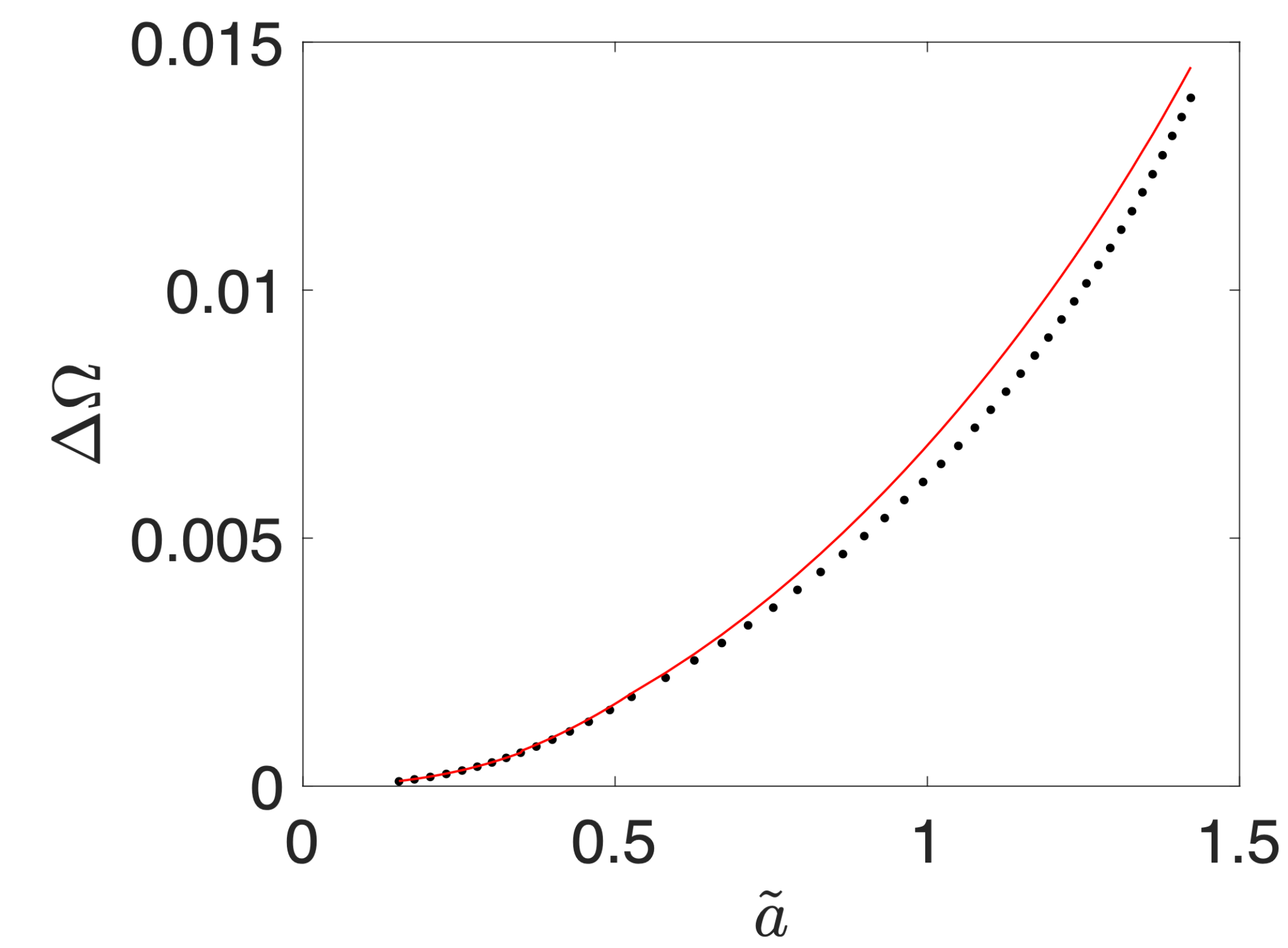}
    \caption{ Traveling breather frequency shift from computation
      (black dotted line) and predicted (red solid line) traveling breathers.  }
    \label{fig_NLS_freq_shift}
\end{figure}

Performing $c$ continuation, we observe an increase in traveling
breather amplitude with decreasing $c$.  Define the traveling breather
nonlinear frequency shift
\begin{equation}
  \label{eq:2}
  \Delta \Omega = \Omega_0 - \Omega ,
\end{equation}
where $\Omega_0 = \omega_0-\tilde{k}\partial_{\Tilde{k}}\omega_0$ is
the linear frequency in the co-moving frame.  In
Fig.~\ref{fig_NLS_freq_shift}(a) the prediction
$\Delta\Omega = \epsilon^2/2$ from Eq.~\eqref{Internal-dynamics-BBM}
is compared with the computed frequency shift \eqref{eq:2} for
family 1 across a range of amplitudes.  As expected, there is good
agreement at low amplitudes and deviation at large amplitudes.

\begin{figure}
  \centering
  \includegraphics[width=\linewidth]{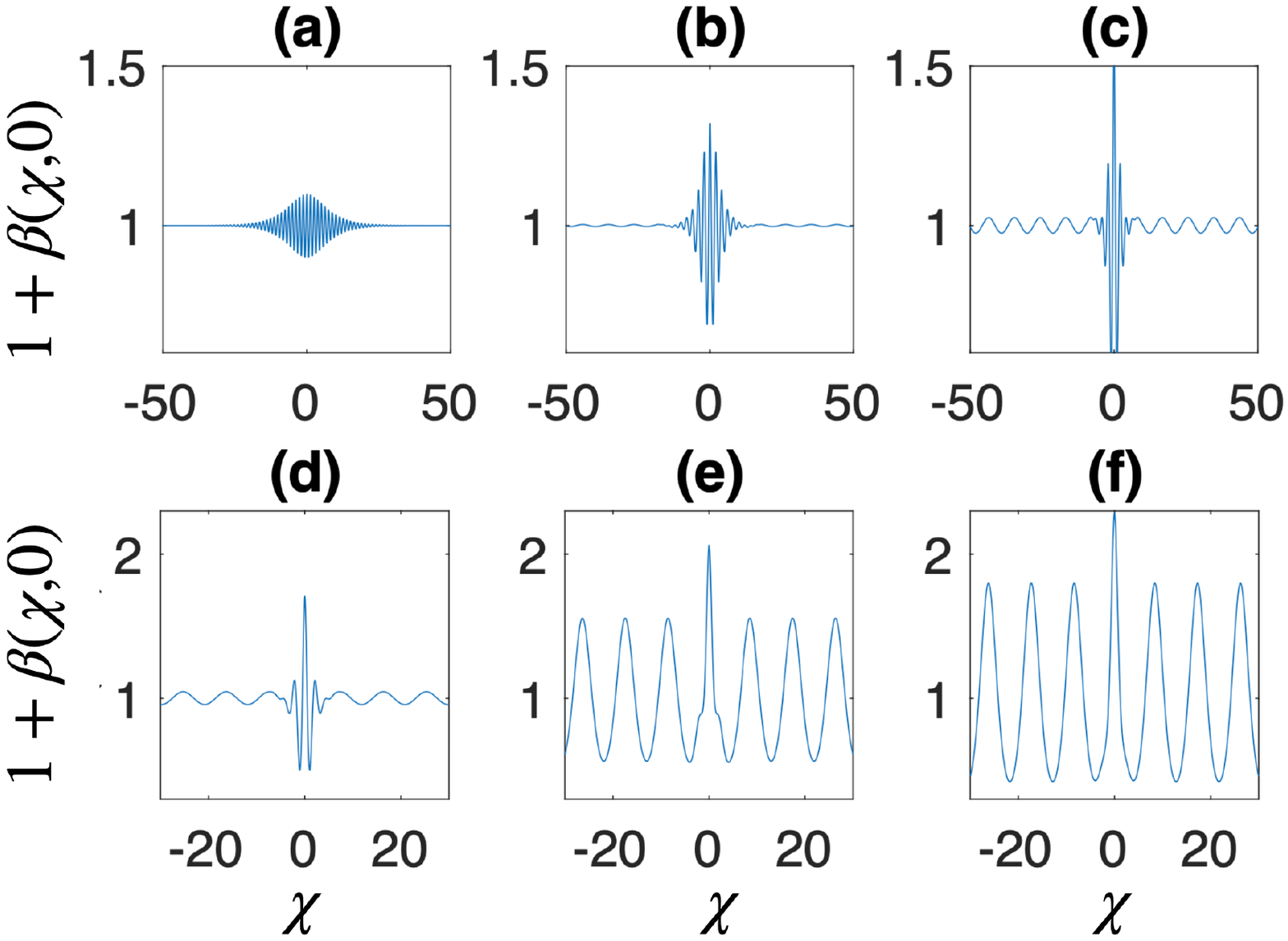}
  \caption{Family 1 of traveling breather solutions of the BBM
    equation with (a) {$(T,c) \approx16.917,-0.036$}, (b) {$(T,c) \approx11.969,-0.076$}, (c) {$(T,c) \approx11.942,-0.079$}, (d) {$(T,c) \approx11.906, -0.082$}, (e){$(T,c) \approx11.316,-0.099$}, (f) {$(T,c) \approx10.682,-0.129$}.}
  \label{fig_BBM_rep_solution_profiles}
\end{figure}

Figure \ref{fig_BBM_rep_solution_profiles} displays six traveling
breather solutions from family 1 at $\tau = 0$.  To the eye, the
solution in Fig.~\ref{fig_BBM_rep_solution_profiles}(a) appears
localized, decaying to 1 as $|\chi| \to \infty$.  In fact, all
computed traveling breathers exhibit an oscillatory background.  This
is consistent with rigorous studies of breathers in Klein-Gordon
equations where it was proven that, in the nonintegrable case, small-amplitude breather solutions are accompanied by exponentially small
oscillatory tails
\cite{segur_nonexistence_1987,soffer_resonances_1999-1}.  As the
traveling breather amplitude increases in
Fig.~\ref{fig_BBM_rep_solution_profiles}, the oscillatory background
becomes more prominent, eventually reaching an amplitude that is
comparable to the traveling breather itself.  The increase in carrier
amplitude is accompanied by a narrowing of the traveling breather's
width such that, in Figs.~\ref{fig_BBM_rep_solution_profiles}(e,f),
the traveling breather itself is narrower than the carrier
wavelength.

We point to some other general trends within the computed solution
families. The upper limit to the traveling breather velocity $c$ is
approached in the vanishing amplitude regime so that it is the linear
group velocity for family
$c < \partial_{\tilde{k}} \omega_0(\tilde{k}^{(j)},1)$, implying that
all traveling breather velocities are negative. {On the other hand,
  the computations do not apparently indicate a lower bound to the breather velocities:}
\begin{equation}
  \label{Linear-group-speedBBM}
 { c <
  \partial_{\tilde{k}} 
  \omega_0(\tilde{k}^{(j)},1) < 0 .}
\end{equation}
This implies that, when bifurcating from the NLS bright soliton,
traveling breather velocities {are always negative, i.e. $c\in (-\infty,0)$.} 

The traveling breather envelope amplitude is defined as
\begin{equation}
  \label{eq:5}
  \tilde{a} = \max_{\tau \in [0,T]} \beta(0,\tau) - \min_{\tau \in
    [0,T]} \beta(0,\tau).
\end{equation}
In
Fig.~\ref{fig_BBM_rep_solution_profiles} we observe that the traveling breather width narrows relative to the
carrier wavelength. 

Let us count the number of parameters characterizing the traveling
breather solutions.  In addition to the velocity $c$ and time period $T$,
the existence of the periodic background introduces two additional
parameters, the carrier amplitude
\begin{equation}
  \label{eq:6}
  a = \limsup_{\chi \to \infty} \beta(\chi,0) -
  \liminf_{\chi \to \infty} \beta(\chi,0) ,
\end{equation}
and the carrier wavenumber $k$ [cf.~Eq.~\eqref{eq:9}], with the
carrier frequency $\omega = \omega(k,a)$ uniquely determined by the
unit-mean ($\overline{u}=1$) constraint \eqref{eq:8}.  Since the
traveling breather frequency in the comoving frame is known
$\Omega = 2\pi/T$, we have the additional compatibility
relation
\begin{equation}
  \label{eq:3}
  \omega(k,a) - ck = \Omega .
\end{equation}
Traveling breathers impart a phase shift to the periodic background,
which we normalize to $\sigma \in [-\pi,\pi]$. The phase shift
quantifies the amount by which the periodic background has
advanced or receded across the traveling breather core.  In our
computations, the phase shift is implicitly determined by the spatial
domain $[-L,L]$ and the imposition of spatially periodic boundary
conditions.  The number of carrier wavelengths that fit in the domain
is $N = \lfloor Lk/\pi \rfloor$ so  the difference
$\Delta \chi = 2L - 2\pi N/k \ge 0$ represents the spatial mismatch.
The normalized phase shift is determined according to
\begin{equation}
  \label{eq:4}
  \sigma =
  \begin{cases}
    k \Delta \chi, & k\Delta \chi < \pi, \\
    k \Delta \chi - 2\pi, & k\Delta \chi > \pi .
  \end{cases}
\end{equation}
For our traveling breather computations, we input the four parameters
$c,T,L,\overline{u}=1$.  The two relations \eqref{eq:3} and
\eqref{eq:4} can be used to determine $a$ and
$\sigma$.  Thus, unit-mean traveling breathers constitute
a three-parameter family of solutions.



\begin{figure}
    \centering
    \includegraphics[width=\linewidth]{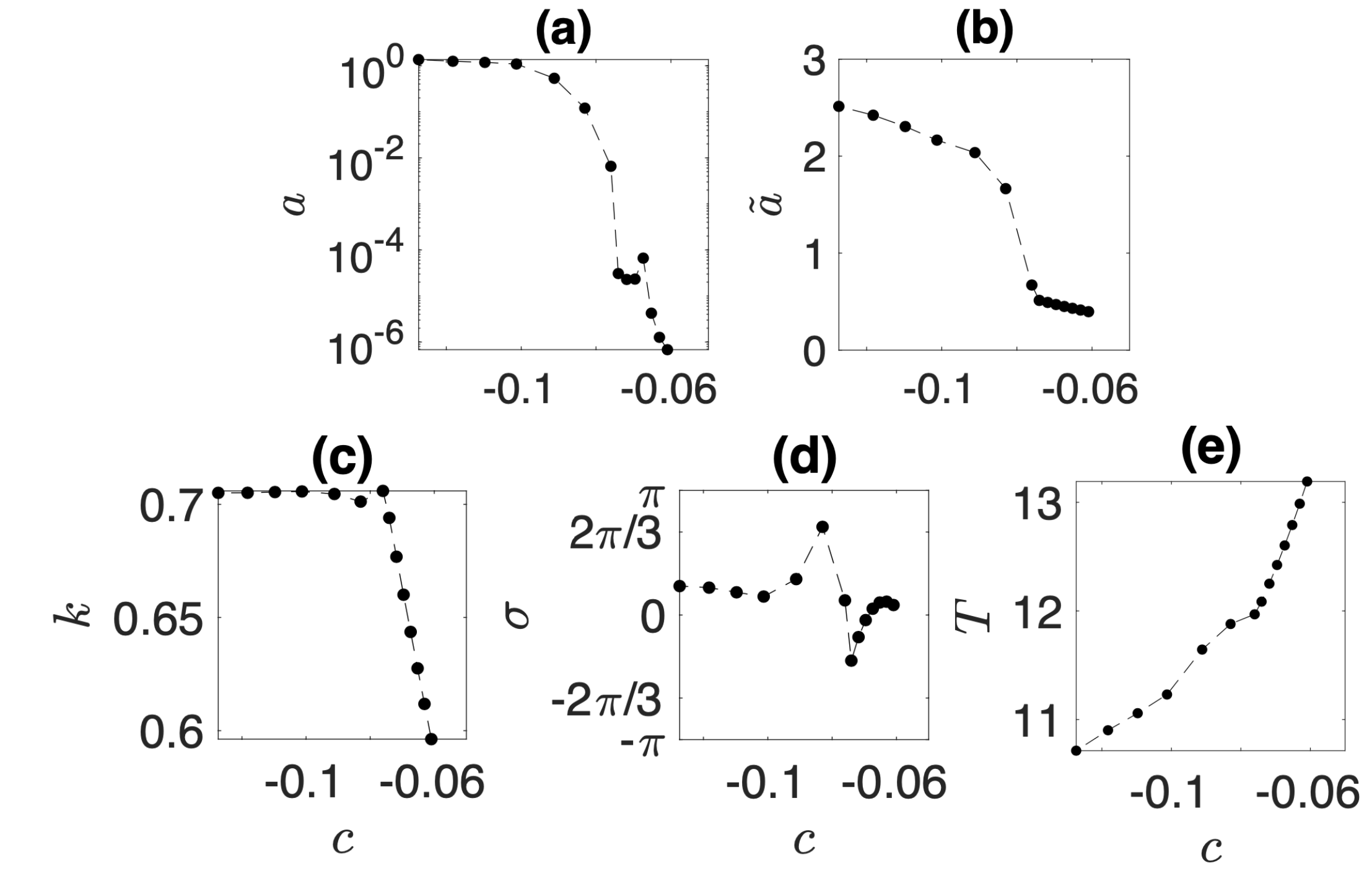}
    \caption{Variation of the (a) periodic background amplitude
      \eqref{eq:5}, (b) breather amplitude \eqref{eq:6}, (c) periodic
      background wave number, (d) phase shift $\sigma$ \eqref{eq:4}, and (e) timeperiod $T$
      with breather velocity $c$ for family 1 of BBM
      breathers.}
    \label{fig:BBM-family-parameterization}
\end{figure}
In Fig.~\ref{fig:BBM-family-parameterization} we show how the
parameters $a$, $\tilde{a}$, $k$ , $\sigma$, and $T$ vary with the
breather velocity $c$ for family 1. As $c$ is decreased, the parameters
$a$, $\tilde{a}$, and $k$ increase, with {$k$ and $a$} rapidly approaching an asymptotic value in the strongly
nonlinear regime. The phase shift {$\sigma$ also limits} to an
asymptotic value for large-amplitude (more negative $c$) traveling
breathers, while displaying both positive and negative phase shifts
across the range of velocities. By continuity, there is a traveling
breather exhibiting a zero phase shift. Finally, we remark that, since
we initialized the computations of family 1 with the NLS bright
soliton \eqref{Internal-dynamics-BBM} where
{$\tilde{a} \approx 0.15$}, the largest value of $c$,
{$c_{\rm max} \approx -0.031$}, is less than the theoretical upper
bound \eqref{Linear-group-speedBBM}
$\partial_{\tilde{k}} \omega_0(\tilde{k}^{({1})},1) \approx
-0.0308$, which applies in the limit $\tilde{a} \to 0$. We did not
attempt to compute smaller-amplitude solutions.  The smallest velocity
for which we compute a traveling breather solution is
$c \approx -0.129 $, which is shown in
Fig.~\ref{fig_BBM_rep_solution_profiles}(f). At this point, the
existence of a lower bound for breather velocities is unknown but
remains interesting for future investigations.

\begin{figure}
  \centering
\includegraphics[width=\linewidth]{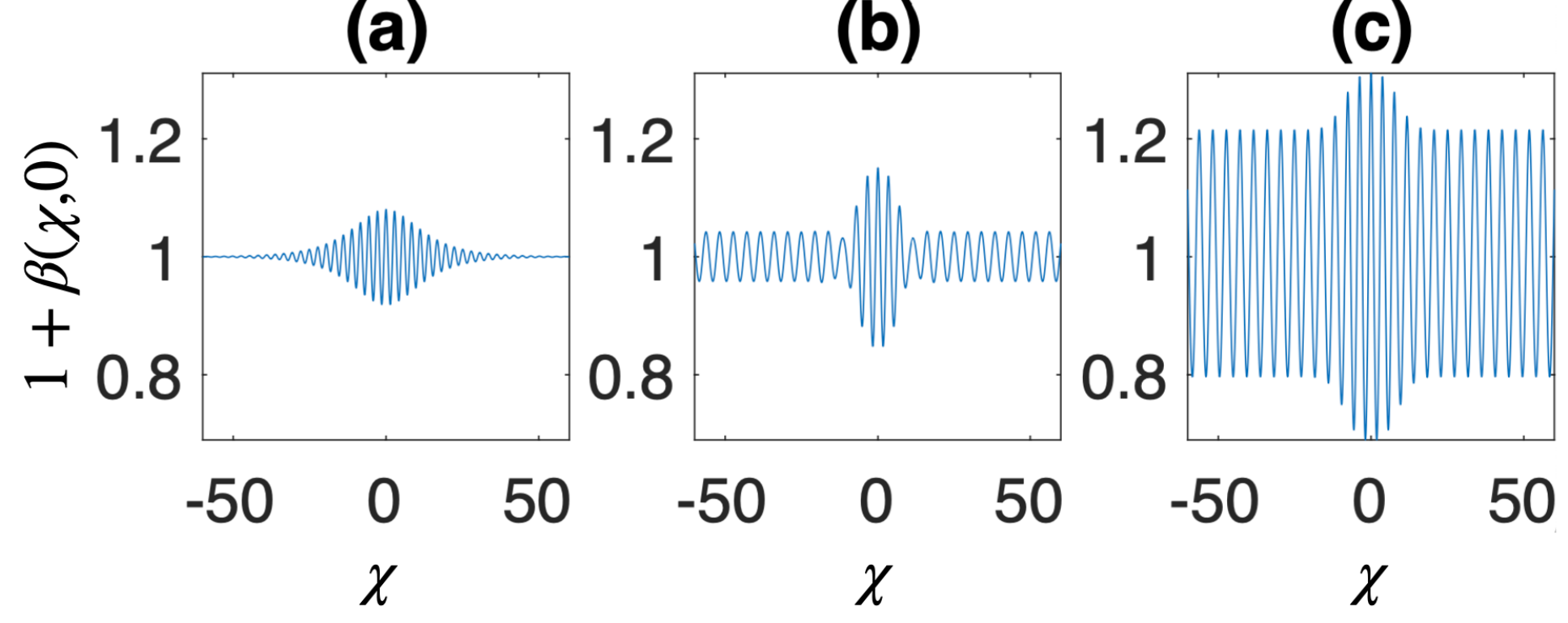}
\caption{Family 5 of BBM traveling breathers with (a) $(\tilde{a},c,T) \approx 0.162,-0.110,10.142$, (b)
  $(\tilde{a},c,T) \approx 0.306,-0.124,9.708$, and (c) $(\tilde{a},c,T) \approx 0.623, -0.133, 9.486$.}
\label{fig_BBM_foc_to_defocus}
\end{figure}
\begin{figure}
    \centering
    \includegraphics[width=1.0\linewidth]{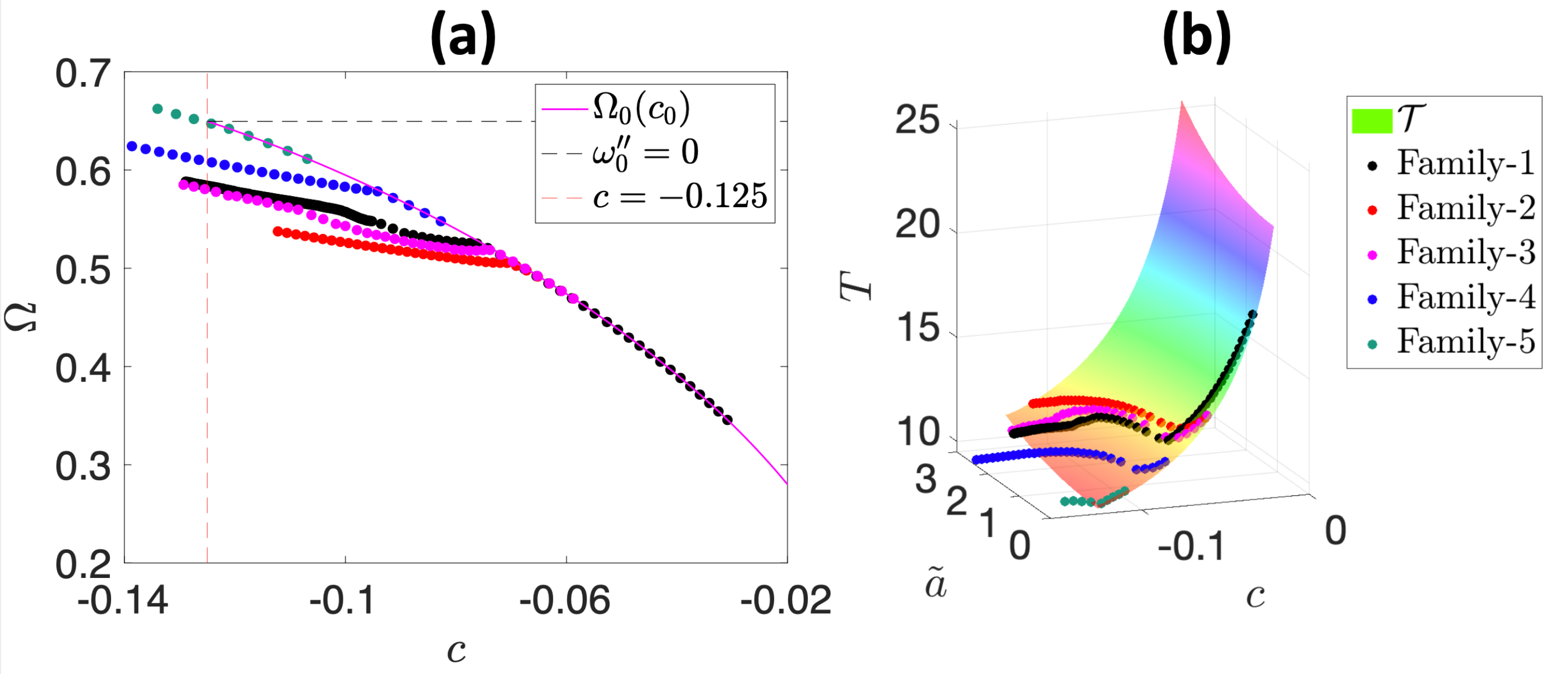}
    \caption{(a) Projections in amplitude of the velocity
      ($c$)-frequency ($\Omega$) relation of the five traveling
      breather branches.  (b) Computed unit-mean traveling breathers from the five
      families (closed circles) and the weakly nonlinear prediction
      $T = \mathcal{T}$ \eqref{NLS-approx-BBM-time-period} (colored
      surface).}
    \label{Bifurcation-diagram-BBM-3D}
\end{figure}
The fifth branch of computed traveling breathers crossing the linear
inflection point shown in Figs.~\ref{fig_BBM_foc_to_defocus} (a)-\ref{fig_BBM_foc_to_defocus}(c)
merits special mention. Given their proximity to the zero-dispersion
line ($\Omega_0 = 3\sqrt{3}/8$) in the weakly nonlinear regime, they
display a prominent periodic background even for small amplitudes. For
amplitudes $0.16\leq\tilde{a}\leq 0.62$, the carrier wavenumber of the
breather core is found to lie in the interval
$\tilde{k} \in [1.67,2.42]$, which is below the linear dispersion
inflection point $\tilde{k} = \sqrt{3} \approx 1.73$.  When
$\tilde{k} < \sqrt{3}$, the NLS equation \eqref{BBM-HNLS} is
repulsive or defocusing.  This persistence of bright traveling breathers
across the zero-dispersion line into the defocusing regime is an
intriguing feature, suggesting the need for a higher-order NLS model
to describe them {\cite{kivshar1991dark,kivshar1998dark}}.

In order to assess how close to the NLS regime computed traveling
breathers are, we plot the relationship between the traveling breather
frequency in the comoving frame with frequency $\Omega$ and velocity
$c$ for each solution family in
Fig.~\ref{Bifurcation-diagram-BBM-3D}(a). To compare with the NLS
bright soliton \eqref{Internal-dynamics-BBM}, we obtain a relationship
between the linear frequency in the co-moving frame
$\Omega_0 = 2\tilde{k}^3/(1+\tilde{k}^2)^2$ and the linear group
velocity
$c_0 = \partial_{\Tilde{k}}\omega_0=
(1-\tilde{k}^2)/(1+\tilde{k}^2)^2$ by eliminating $\tilde{k}$ to
obtain
\begin{equation}
  \label{Omega_func_Omegap}  \Omega_0(c_0) = {2\sqrt{-2c_0}\frac{[(2c_0+1)+\sqrt{8c_0+1}]^{3/2}}{(1+\sqrt{8c_0+1})^2}}, 
\end{equation}
where $-\frac{1}{8} < c_0 < 0$ in the focusing regime.  The computed traveling
breathers lie close to the $\Omega_0(c_0)$ curve in
Fig.~\ref{Bifurcation-diagram-BBM-3D}(a) {for small amplitudes 
  and depart from the curve for larger amplitudes or fewer carrier
  wave oscillations in the breather core.}
{We conclude this section with a concise representation of all the
computed traveling breather families in
Fig.~\ref{Bifurcation-diagram-BBM-3D} (b).} To this end, a convenient set
of defining parameters is the traveling breather amplitude $\tilde{a}$,
the velocity $c$, and the time period in the comoving frame $T$ ({or the associated angular frequency $\Omega=2\pi/T$}). From
Eqs.~\eqref{Internal-dynamics-BBM} and \eqref{eq:2} with
$\Delta \Omega = \epsilon^2/2$, the timeperiod $T$ in the NLS
approximation is
\begin{equation}
  \label{NLS-approx-BBM-time-period}
  \mathcal{T} (\tilde{a},c) = {\frac{2\pi}{\Omega} = \frac{2\pi} {\Omega_0(c) -
    {\frac{\tilde{a}^2} {32}} \gamma(c)}},
\end{equation}
where {$\gamma(c) =\frac{\sqrt{-2c}}{6\sqrt{(2c+1)+\sqrt{8c+1}}}\frac{-5-4c-5\sqrt{8c+1}}{4c-1-\sqrt{8c+1}}$} and
$\tilde{k}(c) \equiv \sqrt{\frac{-(2c+1)-\sqrt{8c+1}}{2c}}$ is
determined by inverting
$c= \partial_{\tilde{k}}\omega_0(\tilde{k},1)$.  All NLS-{like envelope} solitons that {enclose several carrier wave oscillations within the breather core}, lie
close to the two-parameter surface $T = \mathcal{T}(\tilde{a},c)$. {These breather cores decay rapidly to a very small-amplitude ($a\ll 1$) periodic background.}  This surface is depicted in
Fig.~\ref{Bifurcation-diagram-BBM-3D}(b).  The parameters
$(\tilde{a},c,\;{\rm and}\;T)$ associated with each computed traveling breather are
also rendered in the figure.  In the weakly nonlinear regime, the
traveling breathers of all five families reside close to
$T=\mathcal{T}(\tilde{a},c)$.  Eventually, they depart from the
surface, exhibiting a larger time period than weakly nonlinear theory
predicts. {For families $1$-$4$, the strongly nonlinear breathers limit to enclosing very few carrier oscillations within the breather cores, with the envelope widths being comparable to the cnoidal (background) carrier wavelength. An exception in this regard is family 5; wherein the breather core possesses a slowly varying envelope despite a large amplitude $\Tilde{a}$.} {As a final remark, we draw attention to families 2 and 3, computed on domains with $L=100$ and $250$ respectively, and seeded with an identical NLS envelope soliton (see Fig.~\ref{Bifurcation-diagram-BBM-3D}). The role of the computational domain length as an additional parameter (besides $c$ and $T$) is clear, as the scatter plots diverge noticeably beyond the weakly nonlinear regime owing to the different induced phase shift $\sigma$.}
  
\subsection{Conduit equation bright traveling breather solutions}
The examination of BBM bright traveling breathers has primed us for an
investigation of their analogs in the conduit equation, which has an
identical unit-mean linear dispersion relation.

The computations are initialized using the weakly nonlinear NLS
approximation \eqref{Internal-dynamics-BBM} with $\tilde{k} = 4$ and
$\tilde{a} \approx 0.38$. As before, the amplitude scale in
Eq.~\eqref{Internal-dynamics-BBM} is defined by
$\epsilon =\Tilde{a}\gamma/4$, with
$\gamma=(8\Tilde{k}^4+5\Tilde{k}^2+3)/(3\Tilde{k}^5+12\Tilde{k}^3+9\Tilde{k})$. While
continuing the branch of traveling breathers, we observe the familiar
trends associated with a shifting wave mean and the emergence of a
periodic carrier background.  After this family, referred to as
family 1, is computed, we apply the scaling symmetry
\eqref{Scaling-symmetry-conduit-eq} to normalize all the breathers to
unit mean.
The continuation runs slow down once the number of cycles of carrier
wave oscillations in the breather core limit to three, in contrast to
the continuation runs for BBM breathers, where the limit was at one
cycle. Like in the case of BBM breathers, the group velocity
$\partial_{\Tilde{k}}\omega_0=2(1-\tilde{k}^2)/(1+\Tilde{k}^2)^2$ in
the weakly nonlinear regime appears to be an upper limit for conduit
breathers.  This precludes breathers propagating with positive
velocity because $\Tilde{k}>\sqrt{3}>1$. Moreover, a lower limit for breather velocity is not
apparent from the computational results. On the other hand, a point of
difference with the BBM waveforms is a pronounced asymmetry, which
tends to bound the conduit wave profiles away from zero.

\begin{figure}
     \centering
     \includegraphics[width=\linewidth]{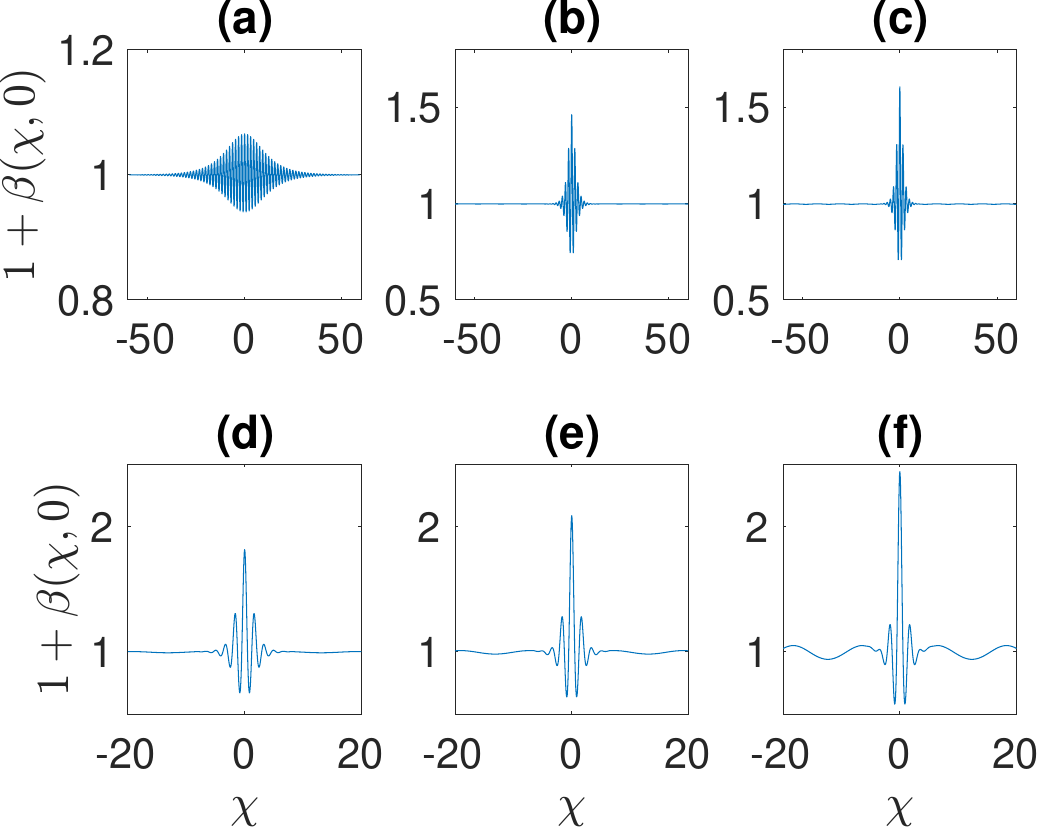}
     \caption{{Family 1 of traveling breathers of the conduit
    equation with (a) {$(T,c) \approx 7.172,-0.101$}, (b) {$(T,c) \approx 7.093,-0.106$}, (c)
    {$(T,c) \approx 7.079,-0.107$}, (d) {$(T,c) \approx 7.073, -0.109$}, (e) {$(T,c) \approx 7.072,-0.111$}, (f) {$(T,c) \approx 7.116,-0.113$}.}}
     \label{fig_conduit_rep_solution_profiles}
 \end{figure}

To illustrate these trends, we present a few representative conduit
breathers from family 1 in
Fig.~\ref{fig_conduit_rep_solution_profiles}. The relevant
computations are performed on a spatial domain of length $2L=120$ with
{spatial step $\Delta x=\frac{2L}{2M}\approx 0.06$ while $N=16$ ($32$ Fourier modes) provides the temporal discretization.} The entire branch is contained within a relatively
small velocity interval of width approximately equal to $0.012$.
 \begin{figure}
    \centering
    \includegraphics[width=1.0\linewidth]{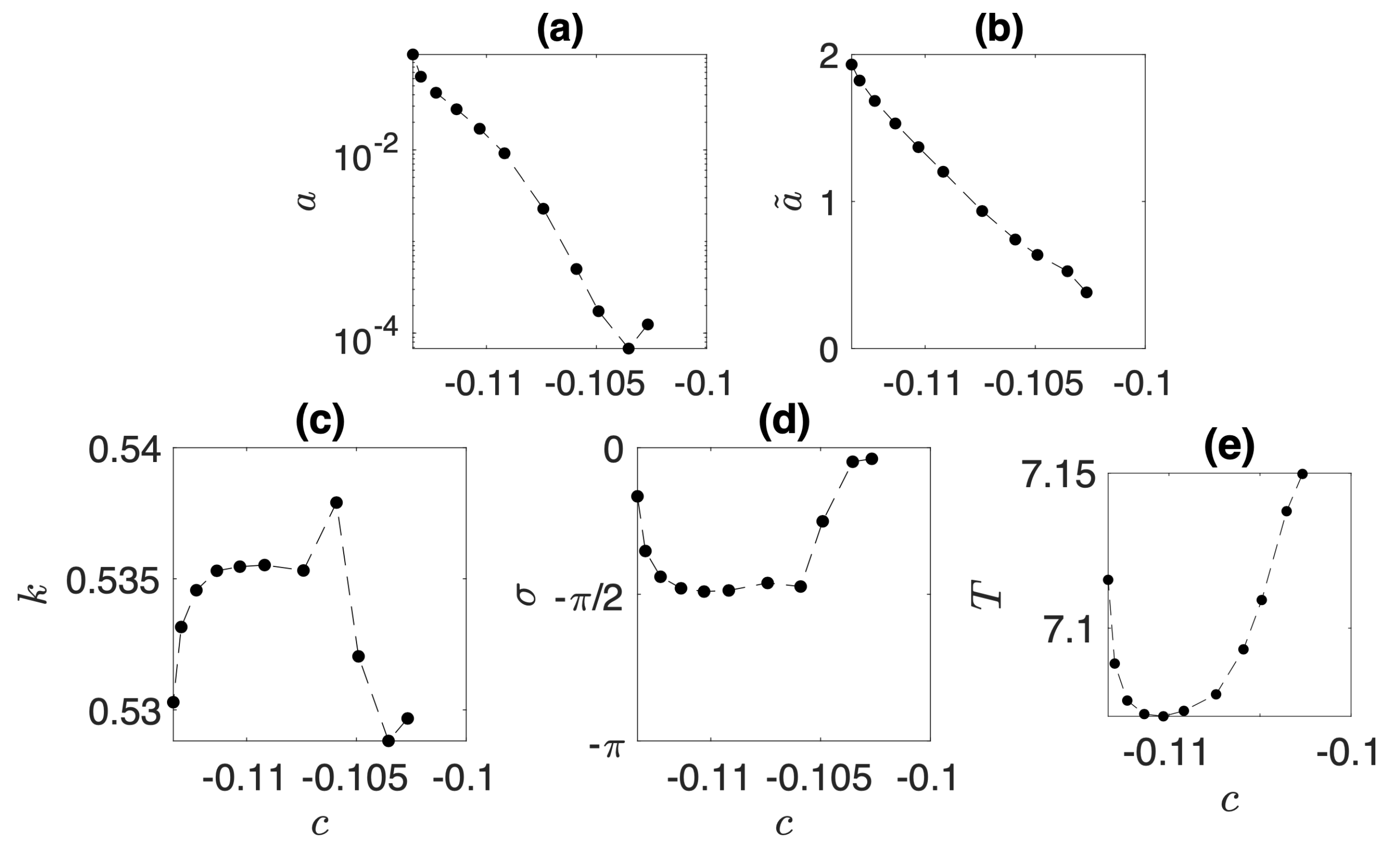}
     \caption{Variation of the (a) periodic background amplitude
      \eqref{eq:5}, (b) breather amplitude \eqref{eq:6}, (c) periodic
      background wave number, (d) phase shift $\sigma$ \eqref{eq:4}, and (e) timeperiod $T$
      with breather velocity $c$ for family 1 of conduit
      breathers.}
    \label{conduit-reparameterization-speed}
\end{figure}
\begin{figure}
     \centering
     \includegraphics[width=\linewidth]{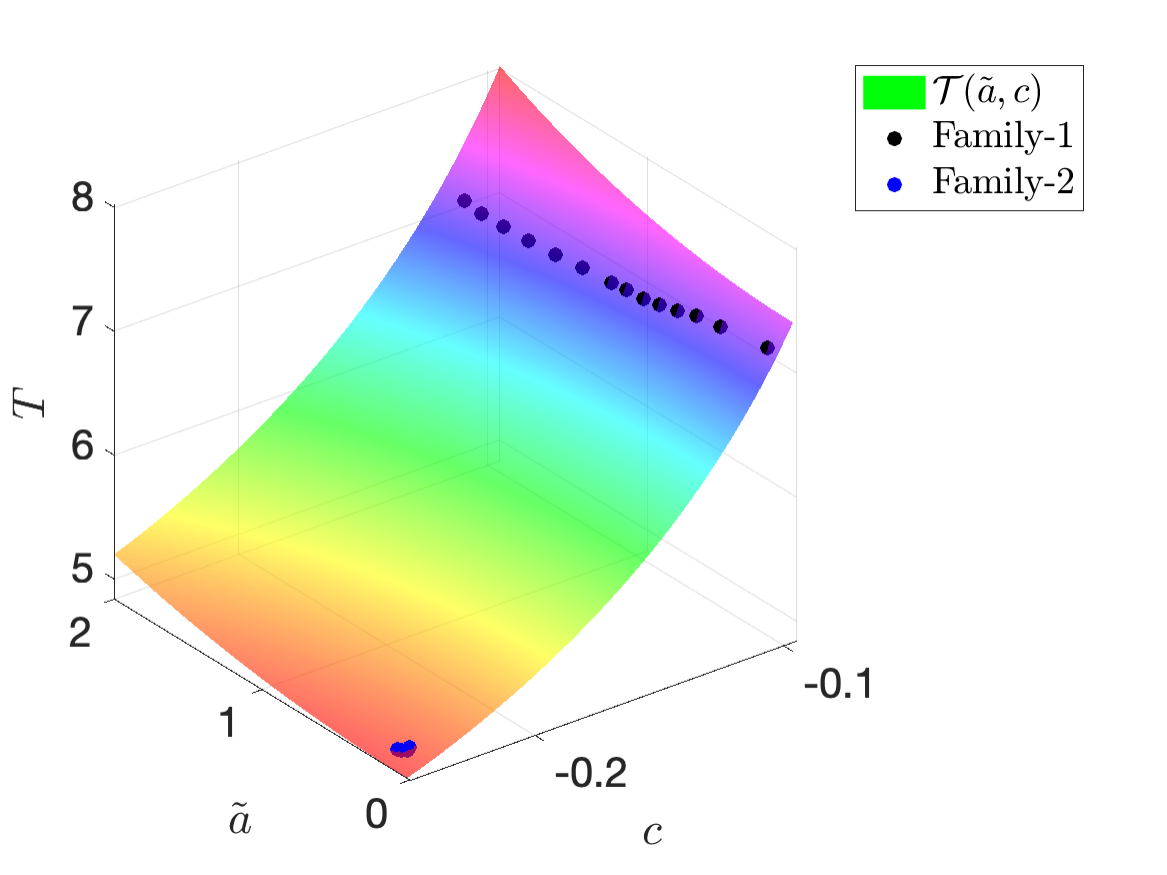}
     \caption{{Computed unit-mean traveling breathers from the 
      families (closed circles) and the weakly nonlinear prediction
      $T = \mathcal{T}$ \eqref{NLS-approx-BBM-time-period} (colored
      surface). }}
     \label{Conduit-breathers-bifurcation-diagram}
 \end{figure}
\begin{figure}
  \centering
  \includegraphics[width=\linewidth]{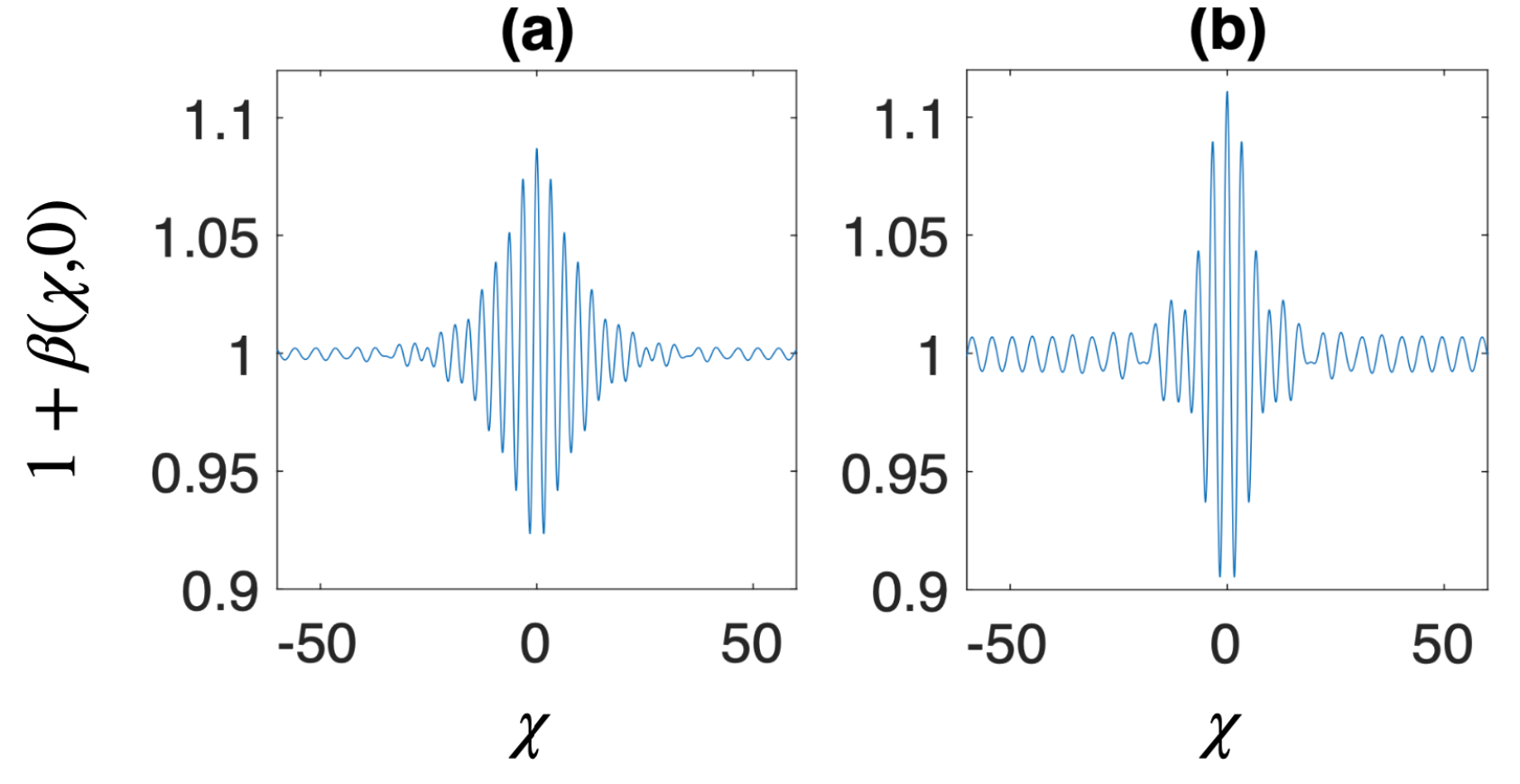}
  \caption{{Family 2 of conduit weakly nonlinear wavepackets near the
      zero-dispersion line, characterized by (a) carrier wave number
      $\tilde k\approx 2$ and
      $(\tilde{a},c,T)\approx 0.166,-0.240,4.907$ and (b)
      $\tilde k\approx 1.8$ and
      $(\tilde{a},c,T)\approx 0.210, -0.243,4.886$.}}
     \label{fig_conduit_rep_solution_profiles-zero-disp}
\end{figure}

In Fig.~\ref{conduit-reparameterization-speed} we show how the
identifying parameters ${a}$, $\tilde{a}$, ${k}$, $\sigma$, and $T$
vary with the breather velocity $c$ for family 1, upon the emergence
of the periodic background. With an increasing magnitude of the
wave-packet speed, increasing trends in ${a}$ and $\tilde{a}$ are
observed, while ${k}$ displays slight variation. The phase jump
$\sigma$ is negative for all breathers in this branch. Given the
strongly nonlinear nature of the conduit equation, it is interesting
to check how close the traveling breathers are to the NLS regime. To
this end, as before, the timeperiod $T$, velocity $c$, and amplitude
$\Tilde{a}$ form a set of identifying parameters. Moreover, the
timeperiod in the NLS approximation is given in
Eq.~\eqref{NLS-approx-BBM-time-period}, where, for the conduit
equation, the linear frequency $\Omega_0(c)$ is
\begin{align}
  &\Omega_0(c)=\frac{2\sqrt{-c}(1+c+\sqrt{4c+1})^{3/2}}{2c+1+\sqrt{4c+1}},
\end{align}
thus necessitating $-\frac{1}{4}<c_0<0$ for a real frequency. Additionally,
$\gamma(c)$ is defined to be
\begin{align}
  &\gamma(c)=\frac{\sqrt{-c}}{3\sqrt{c+1+\sqrt{4c+1}}}\\\nonumber&\times \frac{16+16\sqrt{4c+1}+43c+11c\sqrt{4c+1}+6c^2}{2+2\sqrt{4c+1}+2c-2c\sqrt{4c+1}}.
\end{align}
The NLS surface for the conduit equation $\mathcal{T}(\Tilde{a},c)$ is
shown in Fig.~\ref{Conduit-breathers-bifurcation-diagram}. For
appreciably small intervals in velocity, timeperiod and breather
amplitudes, the wave packets reside on the $\mathcal{T}(\tilde{a},c)$
surface and thereafter lie entirely below it.

We also investigate the existence and form of weakly nonlinear conduit
breathers near the zero-dispersion line.  To this end, we initiate the
computations with an appropriate NLS initial guess.  We recover a
delocalized wavepacket with a nearly monochromatic amplitude-modulated
core with $\tilde{k}=2$ [see
Fig.~\ref{fig_conduit_rep_solution_profiles-zero-disp}.a]. We seed the
$c$-continuation algorithm with this wavepacket, to obtain the
breather in Fig.~\ref{fig_conduit_rep_solution_profiles-zero-disp}(b). This wavepacket is characterized by core carrier wave
oscillations with wavenumber $\tilde{k}=1.8$.  The continuation
procedure is seen to slow down significantly thereafter. It is
remarkable that even for such small-amplitude traveling breathers,
there is a relatively large-amplitude periodic background, which
points to the operable higher-order dispersive effects therein. At
this point, the existence of bright wave packets across the
zero-dispersion line is unclear but is interesting for future
investigation.

The strongly nonlinear nature of the conduit equation, coupled with
the large conditioning numbers of the symmetrized system of linear
equations at each Newton step, result in reduced computational
tractability of the continuation algorithm.  

\subsection{Dynamic stability of breathers}
\label{Dynamic-stability}

We numerically investigate the dynamic stability of computed BBM and
conduit traveling breathers with direct numerical simulations. The
initial condition consists of a numerically computed traveling
breather solution $1 + \beta(\chi,\tau)$ evaluated at $\tau = 0$ that
is multiplicatively perturbed: $u(x,0)=[1+\beta(x,0)][1+\Delta(x)]$
for the BBM equation and similarly for the conduit equation. A smooth
perturbation function $\Delta$ is constructed from a spatially
periodic disturbance with random Fourier series coefficients that is
band limited and scaled to have a peak amplitude of $0.05$. We employ the standard fourth-order Runge-Kutta
explicit time-stepping scheme along with a Fourier discretization in
space similar to \cite{maiden2016modulations,congy2020dispersive}. We
perform long-time numerical integration for more than 100 breather
periods $T$ of two perturbed traveling breather solutions, one weakly
nonlinear and the other strongly nonlinear, in each of the seven
computed wave families (five BBM and two conduit families).  All
exhibit similar dynamical behavior.  A representative example of a
numerically evolved, perturbed strongly nonlinear breather solution
from family 1 of the conduit equation compared with the unperturbed
breather is shown in Figs.~\ref{fig:conduit_breather_dynamic}(a) and (b).  The
traveling breather core is slightly delayed after an evolution time
of $150 T$.  Despite the 5\% initial perturbation and long evolution
time, the breather retains its coherence.

\begin{figure}
    \centering
    \includegraphics[width=\linewidth]{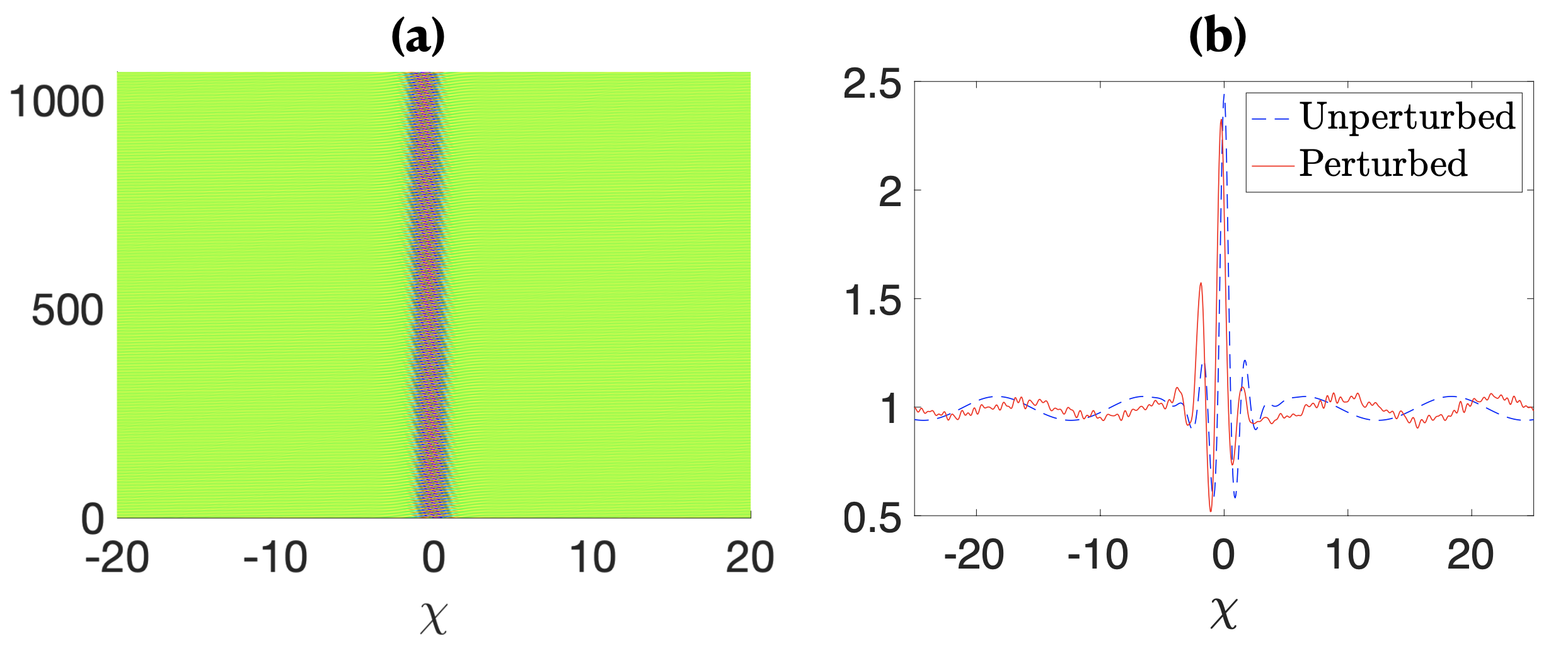}
    \caption{Evolution of a perturbed strongly nonlinear conduit
      traveling breather from family 1. {(a)
        Contour plot describing the spatiotemporal structure of the
        perturbed waveform when evolved to $t=150 T\approx 1067$. (b)
        Time snapshots of the perturbed (red solid line) and unperturbed
        waveforms (blue dashed line) at the end time, pointing to
        the coherence of the breather.  }}
    \label{fig:conduit_breather_dynamic}
\end{figure}
\section{Discussion}
 
\label{Conclusions and discussions}
Branches of bright traveling breathers have been computed as solutions
to a space-time boundary-value problem for the BBM and conduit
equations. For both we found that traveling breathers are
approximated by NLS envelope bright solitons for small amplitudes and
carrier wave numbers sufficiently deep in the negative dispersion
regime.
An emergent feature of traveling breathers in the weakly nonlinear
regime was delocalization, signified by the presence of a
propagating periodic background. While the NLS approximation remains
accurate in the traveling breather core, it is necessary to examine
the effects of higher-order dispersive corrections to explain the
delocalization, which we briefly describe now.

A convenient framework to understand delocalization is the
third-order NLS (TNLS) equation
\begin{align}
  \label{Scaled-NLS}
  &i B_{{\tau}}+\frac{B_{{\zeta}{\zeta}}}{2}+|B|^2 B=
    i\epsilon^{\prime}B_{{\zeta}{\zeta}{\zeta}},0< \epsilon^{\prime}
    \ll  1 ,
\end{align}
where
$\epsilon^{\prime}={\epsilon
}({\partial_{\tilde{k}\tilde{k}\tilde{k}}\omega_0})/{6(\partial_{\tilde{k}\tilde{k}}\omega_0)^{3/2}}$,
the slow, traveling spatial coordinate $\zeta$ is defined in
Eq.~\eqref{Slow-coordinate-system} and $\epsilon$ is the amplitude
scale [cf.~eq.~\eqref{BBM-HNLS}]. Without loss of generality, we
consider the usual unit amplitude NLS envelope soliton
$B(\zeta,\tau) = {\rm sech}(\zeta)\exp(i\tau/2)$. Linearizing
\eqref{Scaled-NLS} about the soliton and seeking a resonant solution
with the same frequency of the form
$\exp\left[i(\kappa {\zeta}+\frac{{\tau}}{2})\right]$, we obtain the
cubic equation for $\kappa$,
\begin{equation}
  \epsilon^{\prime} \kappa^3+\frac{1}{2}\kappa^2+\frac{1}{2}=0,
\end{equation}
which admits exactly one short-wave solution
$\kappa_0\sim -\frac{1}{2\epsilon^{\prime}}$, $|\kappa_0| \gg 1$ as
$\epsilon^{\prime} \to 0$. 
This resonance with the linear spectrum has two implications: (a) NLS
solitons radiate short waves when subject to weak third-order
dispersion and (b) steady solutions to Eq.~\eqref{Scaled-NLS},
$B({\zeta},{\tau})\equiv A\left( {\zeta}-\epsilon^{\prime}c^{\prime}
  {\tau}\right)\exp{i({\tau}/2)}$ are delocalized. 
Both the unsteady and steady problems are studied under the framework
of exponential asymptotics (see \cite{yang2010nonlinear}), which
yields estimates for the amplitudes of the one-sided short-wave
radiation emitted by NLS solitons and the far-field
(${\zeta}\rightarrow \pm \infty$) tails of the steady, delocalized
waveforms.

An examination of the steady ODE problem reveals that
$A(\xi)$ must be exponentially small in $\epsilon^{\prime}$
as $|\xi| \to \infty$.  The complex-valued profile $A(\xi)$ exhibits a
nonzero phase jump. 
The implications of this perturbative analysis for weakly nonlinear
traveling breathers are that they generically admit a
parametrization 
in terms of the carrier wave number, traveling breather amplitude, and
phase jump, respectively. Moreover, this exponential asymptotic result
corroborates our computational finding of small-amplitude traveling
breathers with nearly localized spatial waveforms.

The wavenumber of the resonant wave in the fast spatial coordinate $x$
is
$\tilde{k}_1\equiv
\tilde{k}+\kappa_0\frac{\epsilon}{\sqrt{\partial_{\tilde{k}\tilde{k}}\omega_0}}\sim\tilde{k}-3(\partial_{\tilde{k}\tilde{k}}\omega_0/\partial_{\tilde{k}\tilde{k}\tilde{k}}\omega_0)$. This
TNLS prediction is viable provided
$|\partial_{\tilde{k}\tilde{k}}\omega_0/\partial_{\tilde{k}\tilde{k}\tilde{k}}\omega_0|\ll1$. We have
compared this wavenumber prediction for the BBM and conduit dispersion
with the computed traveling breather solutions and observed a
significant discrepancy in the interval $\Tilde{k}\in [3,3.5]$ for
nearly monochromatic BBM (or conduit breathers). The relative errors
here were found to be as large as $1600\%$. On the other hand, in the
interval $\Tilde{k}\in [1.74,1.9]$, the relative errors were contained
below $20\%$ and were found to be as low as $6\%$ for
$\Tilde{k}\approx 1.8$. This discrepancy is attributed to the
intricate structure of the dispersion relations
\eqref{BBM-Dispersion-relation} and
\eqref{Conduit-equation-disp-relation} for which
$\partial_{\Tilde{k}\Tilde{k}\tilde{k}}\omega_0$ is zero when
$\tilde{k}=\sqrt{2}+1\approx 2.41$ while $\partial_{\Tilde{k}\Tilde{k}}\omega_0$ is zero at
$\Tilde{k}=\sqrt{3}\approx 1.73$.
We suspect that to complete the characterization of the traveling
breather periodic background, a cubic NLS model incorporating the
{full dispersion} of these nonlocal models may be required.

Another implication of the BBM-conduit nonconvex dispersion relation
is the persistence of traveling bright breathers in the weakly
nonlinear defocusing regime (cf.~Fig.~\ref{fig_BBM_foc_to_defocus}). A
preliminary insight into this persistence can be gained within a TNLS
framework \cite{kivshar1998dark,kivshar1991dark}, where it was shown
that traveling breathers near the zero-dispersion point manifest a
bright or antidark waveform instead.
\section{Conclusion}
In summary, 
we have introduced a direct computational method for traveling bright
breathers of nonlinear dispersive equations.
Multiple families of BBM and conduit equation traveling breathers have
been obtained. In the weakly nonlinear regime, these limit to
amplitude-modulated wave packets that are well approximated by the NLS
equation. In the strongly nonlinear regime, these traveling breathers
were seen to be delocalized, bright modulation defects on
cnoidal-type carrier waves. Large-amplitude BBM breathers were seen to
have more pronounced cnoidal backgrounds than the conduit counterparts
we computed. Our computations indicate that BBM and conduit bright
traveling breathers bifurcating from NLS bright solitons propagate
with negative velocities only {and thus it is required to turn to an alternate setup to what is currently being employed, for their experimental generation \cite{mao2023observation}.}
Finally, BBM and conduit traveling breather solutions were found to be
dynamically stable over the course of long-time numerical evolution of
their initially perturbed waveforms.  A more detailed study
investigating the stability of these traveling breathers using Floquet theory is possible future work. {Another
  interesting problem is the experimental generation of bright
  breather trains and of even a breather gas, from an unstable
  periodic wave \cite{maiden2016modulations}. The latter could also
  form the basis for future investigations in other relevant
  geophysical \cite{whitfield2015wave,helfrich2007decay} or
  short-pulse optical scenarios
  \cite{leblond2013models,costanzino2009solitary}}

{Yet another} extension of the present work is the computation of
bright and dark traveling breathers \cite{hoefer_kdv_2023} which are
generated in the KdV equation via the interaction of solitons and
cnoidal waves \cite{kuznetsov_stability_1975}. {Such classes of
  traveling breathers have been observed experimentally over wide
  amplitude ranges in \cite{mao2023observation} and have been seen to
  exhibit qualitatively similar properties to their asymptotic, KdV
  reductions. An open question is how these solutions relate to the
  bright breathers computed here.  Can soliton-cnoidal wave
  interaction solutions be continued to the bright traveling breather
  solutions obtained here that bifurcate from bright soliton solutions
  of the focusing NLS equation? } Our computational method could help
establish the existence and properties of {such} dark and bright
traveling breathers
in the absence of integrable structure. A tantalizing problem is the
existence of more general two-phase solutions that could also be
explored using a similar computational framework.


 
\begin{acknowledgments}
  Our work reached fruition thanks to the inspiring discussions we had
  with Dr.~Patrick Sprenger and Dr.~Ziad Musslimani, both of whom we
  gratefully acknowledge.
\end{acknowledgments}
\label{Section-5}
\appendix
\section{Validation of the NCG algorithm on mKdV breathers}
The integrable \cite{wadati1973modified},
focusing mKdV equation 
\begin{eqnarray}
  \label{mKdV-equation}
  u_t+3u^2u_x+u_{xxx}=0
\end{eqnarray}
is known to possess bright breathers with closed-form expressions. We
assess the performance of the NCG algorithm on recovering the mKdV
breathers situated on a zero background. The expression for the
two-parameter family of zero-mean mKdV breathers in the envelope
reference frame $u(x,t) = \beta(x - ct,t)$ is
\cite{wadati1973modified}
\begin{eqnarray}
  \begin{aligned}
    \label{mKdV-breather-closed-form}
    &\beta(\chi,\tau;\kappa_1,\kappa_2)\\&=2\sqrt{2}\kappa_1
    {\rm
    sech} (\Theta) \frac{\cos(\Xi)- \frac{\kappa_1}{\kappa_2}
    \sin(\Xi) \tanh(\Theta)}{1+\left( \frac{\kappa_1}{\kappa_2}
    \right)^2 \sin^2(\Xi) {\rm sech}^2 (\Theta)},
  \end{aligned}
\end{eqnarray}
where, without loss of generality, the two solution parameters are
positive $\kappa_{1,2}>0$.  These parameters are related to the
velocity of the envelope $c=\kappa_1^2-3\kappa_2^2$ and the nonlinear
angular frequency of carrier oscillations
$\omega=-\Xi_t=-\kappa_2(\kappa_2^2-3\kappa_1^2)$ in the stationary
reference frame \cite{lamb1980elements}.  Additionally,
$\Xi(\chi,\tau)$ and $\Theta(\chi)$ are
\begin{eqnarray}
  \begin{aligned}
    \Xi(\chi,\tau)&=
                    \kappa_2\left[\chi-2(\kappa_2^2+\kappa_1^2)\tau\right]
    \\\Theta(\chi)&=\kappa_1 \chi. 
  \end{aligned}
\end{eqnarray}

\begin{figure}
  \begin{subfigure}
    \centering
    \includegraphics[width=\linewidth]{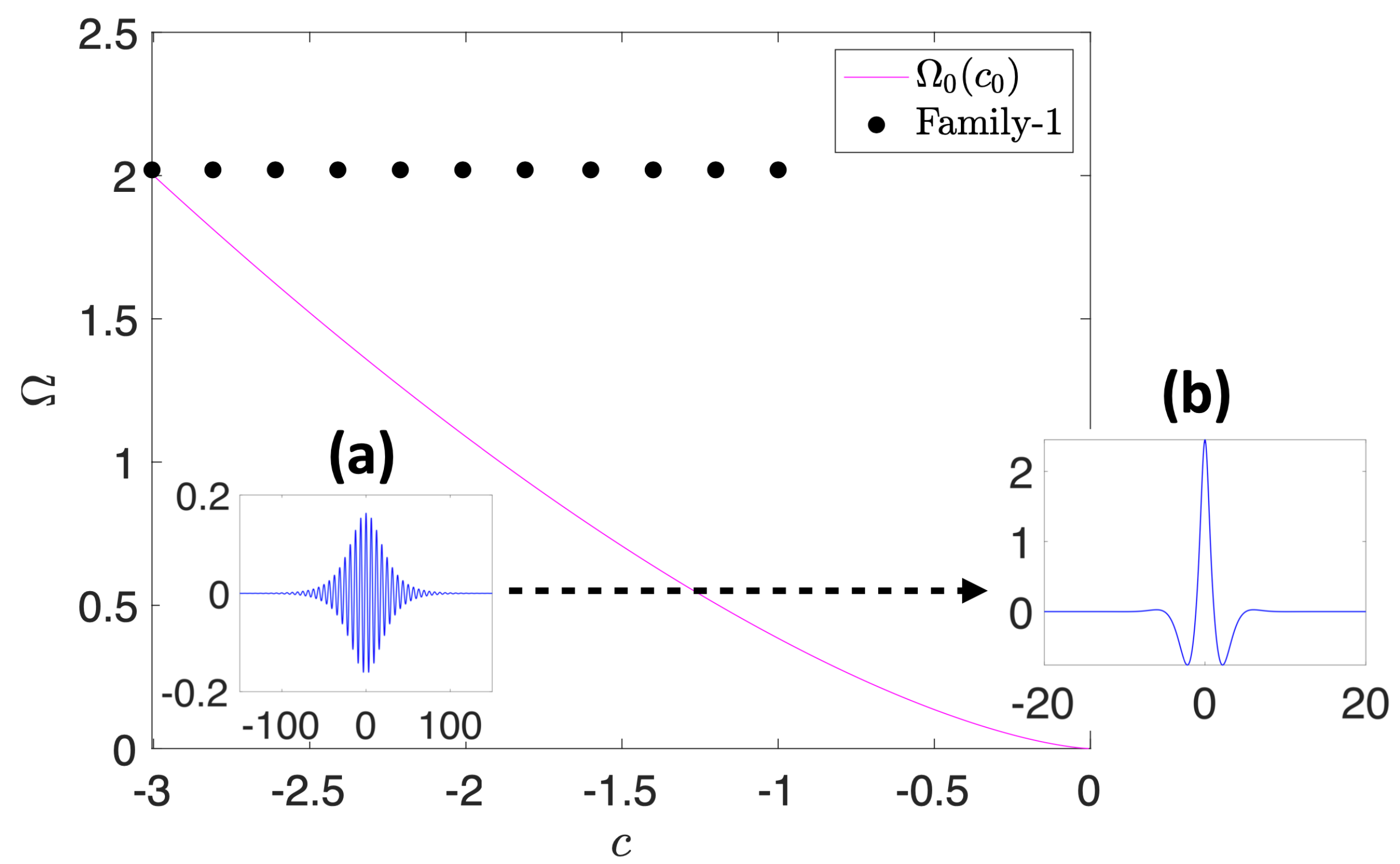}
  \end{subfigure}
  \caption{Continuation path (black dots) marked in the
    $c$-$\Omega$ phase plane; in the weakly nonlinear regime the path
    approaches the linear dispersion curve [magenta curve:
    $\Omega_0(c_0)$] as $c \to -3$. The inset shows slices at $\tau=0$
    of (a) the computed weakly nonlinear mKdV breather and (b) a
    strongly nonlinear breather.}
  \label{fig_mKdV_distinct_breathers}
\end{figure}
There are two distinguished limits of breather solutions to the mKdV
equation (a) the weakly nonlinear NLS regime
(${\kappa_1}/{\kappa_2} \ll 1$) and (b) the strongly nonlinear regime
[${\kappa_1}/{\kappa_2}\sim {O}(1)$] where breathers tend to
double-pole solutions \cite{wadati1982multiple}.  For
$T =2\pi/\Omega=3.1105$, where $\Omega$ is the angular frequency of
oscillations in the envelope reference frame, we recover both the
distinguished limits using the Newton conjugate gradient algorithm,
coupled to a $c$-continuation line search. Our computational
parameters are $\Delta x=0.05$, $\Delta t\approx 0.1$, and the
computational spatial domain $2L=400$.  Along the entire constant $T$
path on the $c$-$\Omega$ plane, the space-time infinity norm of the
error in the numerical solution is kept approximately $
10^{-7}$. The results of numerical continuation are summarized in
Fig.~\ref{fig_mKdV_distinct_breathers}.
\bibliography{Conduit_equation_PRE}
\end{document}